\documentclass[a4paper,11pt]{article}

\usepackage{jheppub}

\newcommand{\norm}[1]{\left\lVert#1\right\rVert}

\title{Cutkosky representation and direct integration}

\author[a, b]{C. Vergu}
\affiliation[a]{Niels Bohr International Academy,\\
17 Blegdamsvej, 2100 K{\o}benhavn, Denmark}
\affiliation[b]{Institute for Gravitation and the Cosmos, Department of Physics, Pennsylvania State University,\\
University Park, Pennsylvania 16802, USA}

\abstract{We present a new method of direct integration of Feynman integrals based on the Cutkosky representation of the integrals.  In this representation we are able to explicitly compute the integrals which yield square root singularities and leave only the integrals which yield logarithmic singularities, thus making the transcendentality weight manifest.  The method is elementary, algorithmic, does not introduce spurious non-physical singularities and does not require a reduction to a basis of pure integrals.}

\begin{document}
\maketitle
\flushbottom

\section{Introduction}
\label{sec:intro}

This paper grew out of trying to understand two basic facts about Feynman integrals.  The first fact is that a large class of Feynman integrals at $L$ loops and in $D$ dimensions can be written as iterated integrals of length $\lfloor \frac {D L} 2\rfloor$.  This is less than or equal to \emph{half} of the number of integrals in momentum space.  It seems that, outside of a few experts, this fact was not widely appreciated and, in fact, before the work of ref.~\cite{Goncharov:2010jf} we lacked even the language to properly discuss transcendental weight in the Feynman integral literature.\footnote{With the notable exception of multiple zeta values which are purely numerical and have no functional dependence (see ref.~\cite{Kotikov:2004er}) and harmonic polylogarithms (see ref.~\cite{Remiddi:1999ew}).}  It is hard to see who should get the credit of this simple but important observation, but Nima Arkani-Hamed has forcefully made this point to me in a private discussion.  This fact is not only surprising, but actually very useful practically since it reduces in half the number of integrals one needs to perform.

The second basic fact is that the iterated integrals are a sequence of \emph{one-dimensional} integrals.  This also looks very surprising.  How can one rewrite the original integral as a sequence of one-dimensional integrals?  And what should these one-dimensional integration variables be?

A first clue to answering these questions arose from an important dichotomy of singularities, noticed already by Landau in his original paper~\cite{landau1959} on singularities of integrals.  As Landau showed, the singularities broadly divide into two categories: square root\footnote{Or perhaps more generally algebraic.} and logarithmic.  These singularities have very different nature.  Square root singularities are \emph{algebraic} and do not contribute to transcendental weight, while logarithmic singularities are transcendental and do contribute to transcendental weight.

Furthermore, ref.~\cite{Hannesdottir:2021kpd} showed in examples that about half of the singularities are of square root type and half are of logarithmic type.  It is then reasonable to guess that it is only the logarithmic singularities which contribute to the transcendental weight and in turn this gives a way to explain Arkani-Hamed's observation.

But the second question remains.  What should be the one-dimensional integration variables?  A second clue towards answering this come from Cutkosky's work~\cite{Cutkosky:1960sp} where he described discontinuities across branch cuts in terms of cut integrals.  In that reference Cutkosky introduced a new way to write the Feynman integrals, designed with the explicit purpose of making both the singularities and the discontinuities manifest.  This representation, as we will review below in sec.~\ref{sec:cutkosky} and sec.~\ref{sec:cuts_spectral}, is a sequence of one-dimensional integrals potentially followed by a higher-dimensional integral.  As a general rule, when this last integral is non-trivial, an example being that of a non-singular elliptic integral, the Feynman integral can not be written as a polylogarithm (we discuss an example in sec.~\ref{sec:two-loop_sunrise}).

In order to make contact with the iterated integral form of the answer, it should then be possible to explicitly do the integrals producing square root singularities.  This would then constitute an alternative way of performing direct integration to the methods of refs.~\cite{Brown2009, Panzer:2014caa}.

We explicitly show how to do this in a few simple examples.  The method relies on nothing more than basic complex analysis and Cauchy theorem.  As we will show, in order for this to work, there should be only one pair of square root branch points at each step involving square root singularities.  This does indeed happen, sometimes via some non-trivial algebraic identities.  The examples presented in this paper are not meant to challenge the state of the art computations, but rather to showcase how the method works in simple examples.  We hope to tackle more complicated cases in future work.

In sec.~\ref{sec:bubble-3d} we discuss the case of a reducible integral and show that applying the same ideas to this case poses no difficulty.  Therefore, this method, unlike the differential equation method (see ref.~\cite{Henn:2013pwa}), can avoid a potentially expensive integral reduction (see ref.~\cite{Laporta:2000dsw}) step.  Clearly, applying the integration algorithm to each Feynman integral separately is not economical.  Instead, as in the unitarity method (see refs.~\cite{Bern:1994cg, Bern:1994zx}), one can group together all the diagrams which contribute to a given Landau singularity (at a given order in perturbation theory), compute the on-shell state sums, take the internal momenta off-shell and compute the resulting integral.  The full answer can be obtained by merging (not adding!) different contributions which account for all potential singularities.

The integration method we introduce has the following advantages over approaches in Feynman parameter space.  First, it can apply to a large variety of mixed types of $i \epsilon$ conditions (advanced, retarded, Feynman, anti-Feynman) which are sometimes required.  This is so because, unlike in Feynman parameter space, we have several denominators and we can choose contours for each independently.  Second, the momentum space method applies effortlessly to cut integrals, which are more difficult in Feynman parameter language (see ref.~\cite{Britto:2023rig}).  A third advantage is that the on-shell varieties arising in momentum space are typically less singular and do not require as many blow-ups (see refs.~\cite{Bloch:2013tra, Berghoff:2022mqu} for examples of blow-ups required in Feynman parameter space).  A more technical difference is that (the properly compactified) contours of integration in momentum space are not relative homology classes so are easier to deal with.  But most importantly, a huge advantage of our method is that we only need to think about one variable at a time and in principle all the singularities in that variable are visualizable in the complex plane of that variable.

The polylogarithmic integrals are fairly well understood and the next frontier is that of integrals which are not polylogarithmic.  If the on-shell space of the leading singularity has a non-trivial topology, such as that of a (non-singular) elliptic curve or a Calabi-Yau variety, then the integrals can not be computed in terms of polylogarithms.  Sometimes, as in the case of the bubble integral in three dimensions, the on-shell space has the topology of a circle.  When complexified, the circle becomes non-compact and adding two points to achieve a compactification amounts to adding two singularities of pole type.

In refs.~\cite{Abreu:2014cla, Abreu:2021vhb} a formalism for defining a coaction has been developed by using cuts instead of differentials.  In principle this can be applied to integrals of elliptic or Calabi-Yau type.  However, the entries of the ensuing symbol will not be as simple as in the polylogarithmic case.  It is therefore not clear yet how to use this method for writing the answer in a canonical form or how effective this method can be in that case.  Indeed, unlike for the polylogarithmic case, even the notion of a prefactor for the integral does not seem to be well-defined (see ref.~\cite{brown2017notes}).  Other approaches have been proposed in refs.~\cite{Adams:2017ejb, Broedel:2018iwv, Wilhelm:2022wow}.

\section{Cutkosky's argument}
\label{sec:cutkosky}

In ref.~\cite{Cutkosky:1960sp}, Cutkosky described a change of variable from the usual loop momentum integration variables to $q_e^2$, where $q_e$ are internal (not necessarily independent) loop momenta.  One can change variables from $k_i$, independent loop momenta to $q_e^2$ and other ``angular'' (in Cutkosky's terminology) variables $\xi$.

After this change of variables the integral reads
\begin{equation}
    \label{eq:cutkosky-form}
    \int_{a_1}^{b_1} \frac{d q_1^2}{q_1^2 - m_1^2} \dotso \int_{a_m}^{b_m} \frac{d q_m^2}{q_1^2 - m_m^2} \int_\gamma \frac{d \xi_1 \wedge \cdots \wedge d \xi_{n - m}}{J},
\end{equation}
where $J$ is a Jacobian factor.  It can potentially contain numerator factors of the original integral as well.

In favorable cases, the last form $\frac{d \xi_1 \wedge \cdots \wedge d \xi_{n - m}}{J}$ has further residues and its integral can be computed in simple terms.  In more complicated cases, this last form is a holomorphic form on elliptic curves (or hyper-elliptic curves) or Calabi-Yau manifolds and $\gamma$ in eq.~\eqref{eq:cutkosky-form} is a real homology cycle.  Despite much study, a general theory of integrals of elliptic or Calabi-Yau type is not available yet.

Then, Cutkosky describes the integration limits as the solutions to a modified form of Landau equations, $\sum_{j \leq i} \beta_j q_j = 0$ where $q_j$ have norms fixed by the values of the outer integrals.  This relies on some (arbitrary) ordering of the propagators.  Obviously, a judicious choice of ordering can simplify the calculations.

It is worth pointing out that the $q_e^2$ integrals in Cutkosky's representation~\eqref{eq:cutkosky-form} have a superficial resemblance to $G$-functions
\begin{equation}
    \label{eq:G-function}
    \int_0^1 \frac{d t_1}{t_1 - a_1} \int_0^{t_1} \frac{d t_2}{t_2 - a_2} \dotso \int_0^{t_{r - 1}} \frac{d t_r}{t_r - a_r}.
\end{equation}
However, the $G$-function representation is more restricted since the boundaries of integration depend in a much simpler way on the previous integration variables.  In Cutkosky's representation the integration boundaries have a potentially complicated functional dependence on previous integration variables $a_s(q_1^2, \dotsc, q_{s - 1}^2)$ and $b_s(q_1^2, \dotsc, q_{s - 1}^2)$ instead.

Nevertheless, Cutkosky's representation has one important qualitative similarity to the $G$-functions: in both cases we are dealing with one-dimensional integrals on Riemann spheres $\mathbf{P}^1$.

In this representation of the integral, the singularities arise as follows (see ref.~\cite[eq.~9]{Cutkosky:1960sp}).  If we denote the result of doing all integrals except the outer one by $F_{(1)}(q_1^2, p)$ where $p$ are external kinematics, then the full answer is
\begin{equation}
  \label{eq:cutkosky-integral}
    \int_{a_1}^{b_1} \frac{d q_1^2}{q_1^2 - m_1^2} F_{(1)}(q_1^2, p).
\end{equation}
If there is a singularity when $p \to p_0$, then this means that the integration contour in the $q_1^2$ complex plane is pinched between $q_1^2 = m_1^2$ and a singularity of $F_{(1)}$.  More precisely, we must have that $F_{(1)}(m_1^2, p)$ is singular when $p \to p_0$.  By a contour deformation we can pick up a residue at $q_1^2 = m_1^2$ and this is the only part of the integral which is singular.  Therefore, the singularity is given by $\pm 2 \pi i F_{(1)}(m_1^2, p)$.  As remarked above, this indeed becomes singular when $p \to p_0$.

\section{Cuts and spectral densities}
\label{sec:cuts_spectral}

Given a function $\rho(x)$, we can build a function $F(x)$ with a branch cut between $a$ and $b$ whose discontinuity is $\rho(x)$.  This construction is well-known and
\begin{equation}
  \label{eq:spectral-representation}
    F(z) = \frac 1 {2 \pi i} \int_a^b \frac{d x}{x - z} \rho(x).
\end{equation}
Indeed, we have
\begin{multline}
    F(z + i \epsilon) - F(z - i \epsilon) =
    \frac 1 {2 \pi i} \int_a^b d x \Bigl(\frac{1}{x - z - i \epsilon} - \frac{1}{x - z + i \epsilon}\Bigr) \rho(x) = \\ =
    \begin{cases}
        \rho(z), & \qquad z \in (a, b), \\
        0, & \qquad \text{otherwise}
    \end{cases}
\end{multline}
where we have used $\frac{1}{x \mp i \epsilon} = \operatorname{pv} \frac{1}{x} \pm i \pi \delta(x)$.

Notice the obvious similarity between eq.~\eqref{eq:cutkosky-integral} and eq.~\eqref{eq:spectral-representation}.  The function $F_{(1)}$ in eq.~\eqref{eq:cutkosky-integral} is the cut of the function defined by the integral.  The function $F_{(1)}$ itself can be represented by a similar integral and the function $F_{(2)}$ it contains corresponds to another cut, where more propagators are set on-shell.  In turn, this writing looks similar to the writing in terms of $G$-functions, with one major difference: the number of integrals in the $G$-function representation is equal to $\lfloor \frac {L D} 2\rfloor$, while the number of integrals in the Cutkosky representation is at least the number of propagators.  We will show below that by explicitly integrating square root singularities and the ``angular integrals'' one can make the number of integrals match.

Clearly in the representation of eq.~\eqref{eq:spectral-representation} $z = a$ and $z = b$ are branch points.  If $\rho(a)$ or $\rho(b)$ are non-vanishing finite constants then we have logarithmic branch points.  It is also possible to have singularities such as $\rho(z) \sim (z - a)^\gamma$ when $z \to a$, for $\Re \gamma > -1$ (to ensure convergence) and similarly for $z \to b$.  In this case the value of $\rho$ at the branch points is either zero or infinity, depending on the sign of $\Re \gamma$.

If $\gamma$ is a half-integer then we have a square root branch point at $z = a$.  Obviously, the type of branch cut (i.e.\ square root versus logarithmic) must match at $z = a$ and $z = b$.  A common form for $\rho$ which we will encounter in the following is $\rho(z) = \frac{1}{\sqrt{(z - a)(b - z)}}$ or more generally $\rho(z) = \frac{1}{\sqrt{(z - a)(b - z)}} P(z)$, where $P(z)$ is a polylogarithmic function.

This is closely connected to the Mandelstam representation (see ref.~\cite{PhysRev.115.1741}).  The Mandelstam representation has been the subject of many studies (see also ref.~\cite{Caron-Huot:2014lda} for an application in a similar context to the present one).  Our proposal amounts to building some kind of spectral densities in perturbation theory.  However, unlike in Mandelstam's approach, we explicitly integrate the square root cuts and keep only the logarithmic singularities.  For example, the double-spectral function for a box diagram is of square root type (see ref.~\cite[eq.~B-42, p.~215]{nakanishi1971graph}) which in our approach would not survive the integration.

\section{The bubble integral}
\label{sec:bubble}

\begin{figure}
    \centering
    \includegraphics{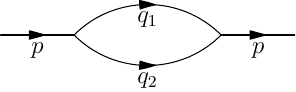}
    \caption{Kinematics of the bubble integral.}
    \label{fig:bubble_integral}
\end{figure}

\begin{figure}
    \centering
    \includegraphics{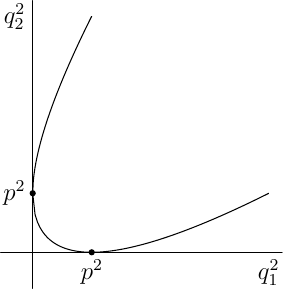}
    \caption{The integration region for the bubble integral in the Euclidean region is inside the curve.  It is outside the curve, including for negative values of $q_1^2$ and $q_2^2$ in the Lorentzian region (for $p^2 > 0$).}
    \label{fig:bubble_integration_region}
\end{figure}

Consider the bubble integral
\begin{equation}
    \int \frac{d^2 q_1}{(q_1^2 - m_1^2) (q_2^2 - m_2^2)},
\end{equation}
where $p = q_1 + q_2$ and $p$ is the external momentum (see fig.~\ref{fig:bubble_integral}).  This can be rewritten as
\begin{equation}
    \int \frac{d q_1^2}{q_1^2 - m_1^2} \int_{(q_2^2)_{\text{min}}}^{(q_2^2)_{\text{max}}} \frac{d q_2^2}{q_2^2 - m_2^2} \int \frac{d^2 q_1}{d q_1^2 \wedge d q_2^2},
\end{equation}
where $(q_2^2)_{\text{min}}$ and $(q_2^2)_{\text{max}}$ are the minimum and maximum values of $q_2^2$, subject to the constraints that $q_1^2$ is fixed and $p = q_1 + q_2$.  The integration domain is in fig.~\ref{fig:bubble_integration_region}.

\subsection{Euclidean signature}
\label{sec:bubble-euclidean}

In Euclidean signature we have
\begin{equation}
    \frac{d^2 q_1}{d q_1^2 \wedge d q_2^2} = -\frac{1}{4 \epsilon(q_1, q_2)},
\end{equation}
where $\epsilon(v, w) = v^0 w^1 - v^1 w^0$.  We have
\begin{equation}
    (q_1 \cdot q_2)^2 + \epsilon(q_1, q_2)^2 = q_1^2 q_2^2.
\end{equation}
Then, the integral to compute becomes
\begin{equation}
    \int_0^\infty \frac{d q_1^2}{q_1^2 + m_1^2} \int_{(\norm{p} - \norm{q_1})^2}^{(\norm{p} + \norm{q_1})^2} \frac{d q_2^2}{q_2^2 + m_2^2} \frac{1}{4 \sqrt{q_1^2 q_2^2 - (q_1 \cdot q_2)^2}},
\end{equation}
where $\norm{p} = \sqrt{p^2}$  and we have used the fact that the minimal value of $q_1^2$ in Euclidean signature is zero and its maximal value is infinity.  Once the value of $q_1^2$ is fixed, the minimal value of $q_2^2$ is $(\norm{p} - \norm{q_1})^2$, obtained when $q_1$ and $p$ are aligned and the maximal value is $(\norm{p} + \norm{q_1})^2$ when $q_1$ and $p$ are anti-aligned.

The inner integral reads
\begin{equation}
    \int_{a_1}^{b_1} \frac{d x_1}{(x_1 + c_1) \sqrt{(b_1 - x_1)(x_1 - a_1)}},
\end{equation}
where $x_1 = q_2^2$, $c_1 = m_2^2$, $b_1 = (\norm{p} + \norm{q_1})^2$ and $a_1 = (\norm{p} - \norm{q_1})^2$.

In general, the integral
\begin{equation}
    \int_a^b \frac{d x}{(x + c) \sqrt{(b - x)(x - a)}}
\end{equation}
with $a < b$ and $c \not\in (a, b)$ can be computed as follows.  We introduce a curve $y^2 = (b - x)(x - a)$ which can be rationally parametrized by $t$ as follows
\begin{gather}
    x = \frac{a + b}{2} + \frac{a - b}{2} \frac{t + t^{-1}}{2}, \\
    y = \frac{a - b}{2} \frac{t - t^{-1}}{2}.
\end{gather}
Then, the integrand can be written as
\begin{equation}
    \omega = \frac{d x}{(x + c)\sqrt{(b - x)(x - a)}} =
    \frac{1}{\sqrt{(a + c)(b + c)}} d \log\Bigl(\frac{t - t_+}{t - t_-}\Bigr).
\end{equation}
The value $x = a$ corresponds to $t = 1$ while $x = b$ corresponds to $t = -1$.  Then we obtain
\begin{equation}
    \int_a^b \frac{d x}{(x + c)\sqrt{(b - x)(x - a)}} =
    \frac{1}{\sqrt{(a + c)(b + c)}} \int_{1}^{-1} d \log \Bigl(\frac{t - t_+}{t - t_-}\Bigr) =
    \frac{\pi}{\sqrt{(a + c)(b + c)}}.
\end{equation}
The logarithm contributes $\pi$, so the transcendental weight is purely numerical and does not have any dependence on the kinematics.

The same integral can be done by contour integration.  Let us briefly describe the method since it will generalize to more complicated cases.  We want to define a function $\sqrt{(b - x)(x - a)}$ to have a branch cut along the segment $[a, b]$.  We pick $b > a$ and $c \not\in [a, b]$.  With the usual definition of the square root for complex numbers we have that $\sqrt{(b - x)(x - a)}$ has a branch cut $(-\infty, a]$ and another branch cut $[b, \infty)$.  Since we want to have a branch cut along the segment $[a, b]$, we split the square root as $\sqrt{(x - a)(x - b)} \to \sqrt{x - a} \sqrt{x - b}$ and we use the definition $\sqrt{z} = \sqrt{\rho} e^{i \frac {\theta} 2}$ where $z = \rho e^{i \theta}$ with $\theta \in [0, 2 \pi)$ for the first square root and the definition $\sqrt{z} = \sqrt{\rho} e^{i \frac {\theta} 2}$ where $z = \rho e^{i \theta}$ with $\theta \in (-\pi, \pi]$ for the second square root.  If we define $z - a = \rho_1 e^{i \theta_1}$ with $\theta_1 \in [0, 2 \pi)$ and $z - b = \rho_2 e^{i \theta_2}$ with $\theta_2 \in (-\pi, \pi]$, then we have that the of the square root above the cut is $i \sqrt{\rho_1 \rho_2}$.  Since for $x \in [a, b]$ we have $b - x = \rho_2$ and $x - a = \rho_1$ have that if defined such that the branch cut is along $[a, b]$, the function $\sqrt{x - b} \sqrt{x - a}$ (where the two square roots are defined as above) is equal by continuity from above the cut to $i \sqrt{(b - x)(x - a)}$.

\begin{figure}
  \centering
  \includegraphics{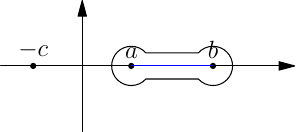}
  \caption{A contour around a square root branch cut.}
  \label{fig:sqrt-cut1}
\end{figure}

Then we have
\begin{equation}
  \int_a^b \frac {d x}{x + c} \frac 1 {\sqrt{(b - x)(x - a)}} =
  \frac i 2 \lim_{\epsilon \to 0} \int_\gamma \frac {d x}{x + c} \frac 1 {\sqrt{x - a} \sqrt{x - b}},
\end{equation}
where $\gamma$ is a contour as in fig.~\ref{fig:sqrt-cut1}.

We have
\begin{equation}
  \int_\gamma \frac {d x}{x + c} \frac 1 {\sqrt{(x - a)(x - b)}} - 2 \pi i \operatorname{Res}_{x = -c} \frac {d x}{x + c} \frac 1 {\sqrt{x - a} \sqrt{x - b}} = 0.
\end{equation}
We therefore have
\begin{equation}
  \label{eq:pole-sqrt-integral}
  \int_a^b \frac {d x}{x + c} \frac 1 {\sqrt{(x - a)(x - b)}} = \frac {\pi}{\sqrt{a + c} \sqrt{b + c}}.
\end{equation}

Finally, we continue with the evaluation of the outer integral
\begin{equation}
    \pi \int_0^\infty \frac{d x_1}{(x_1 + c_1) \sqrt{(x_1 - a_1 + b_1 i)(x_1 - a_1 - b_1 i)}},
\end{equation}
with $x_1 = q_1^2$, $c_1 = m_1^2$, $a_1 + b_1 i = (\norm{p} - i m_2)^2$.  The integral is along the real axis where the quantity under the square root is positive so, with the usual convention for the square root branch cut, the integration path does not intersect the cut.

To compute the integral
\begin{equation}
    \int_0^\infty \frac{d x}{(x + c) \sqrt{(x - a + b i)(x - a - b i)}}
\end{equation}
we introduce a curve $y^2 = (x - a)^2 + b^2$.  This curve can be parametrized rationally by
\begin{gather}
    x = a + b \frac{t - t^{-1}}{2}, \qquad
    y = b \frac{t + t^{-1}}{2}.
\end{gather}
Then we have
\begin{equation}
    \omega = \frac{d x}{(x + c) \sqrt{(x - a + b i)(x - a - b i)}} =
    \frac{1}{\sqrt{(a + c)^2 + b^2}} d \log \frac{t - t_+}{t - t_-},
\end{equation}
where
\begin{equation}
    t_\pm = \frac{-(a + c) \pm \sqrt{(a + c)^2 + b^2}}{b}.
\end{equation}
To finish the computation of the integral we need to compute the values of $t$ at the boundary of the integration region.  At $x = 0$ we have $y = \sqrt{a^2 + b^2}$, where we need to pick the positive sign for $y$.  This implies that
\begin{gather}
    t - t^{-1} = -\frac{2 a}{b}, \qquad
    t + t^{-1} = \frac{2 \sqrt{a^2 + b^2}}{2},
\end{gather}
whence $t_i = \frac{-a + \sqrt{a^2 + b^2}}{b}$.  For $x \to \infty$, we also have $y \to \infty$.  The condition $x \to \infty$ follows from $t \to \infty$ or from $t \to 0^-$, but in the second case we obtain $y \to -\infty$.

Finally, the integral we wanted to compute becomes
\begin{equation}
    \int_{t_i}^\infty \omega =
    -\frac{1}{\sqrt{(a + c)^2 + b^2}} \log \frac{t_i - t_+}{t_i - t_-}.
\end{equation}
Plugging in the values for $a$, $b$ and $c$ we find that the bubble integral in Euclidean signature reads
\begin{equation}
    \frac{\pi}{\sqrt{\Delta}} \log \frac{p^2 + m_1^2 + m_2^2 + \sqrt{\Delta}}{p^2 + m_1^2 + m_2^2 - \sqrt{\Delta}},
\end{equation}
where
\begin{equation}
    \Delta = (p^2)^2 + m_1^4 + m_2^4 + 2 p^2 m_1^2 + 2 p^2 m_2^2 - 2 m_1^2 m_2^2.
\end{equation}

\subsection{Lorentzian signature}
\label{sec:bubble-lorentzian}

In two-dimensional Lorentz signature we have
\begin{equation}
    \frac{d^2 q_1}{d q_1^2 \wedge d q_2^2} =
    \frac{d^2 q_1}{4 (q_1 \cdot d q_1) \wedge (q_2 \cdot d q_2)} =
    \frac{1}{4 \epsilon(q_1, q_2)},
\end{equation}
where we have used $d q_2 = -d q_1$ (since $p$ is considered constant) and fact that $(v \cdot d q_1) \wedge (w \cdot d q_1) = -\epsilon(v, w) d^2 q_1$ with $v \cdot d q = v^0 d q^0 - v^1 d q^1$.  Here we have defined
\begin{equation}
    \epsilon(v, w) = v^0 w^1 - v^1 w^0,
\end{equation}
and $d^2 q = d q^0 \wedge d q^1$.

It can be checked by a simple calculation that
\begin{equation}
    -\epsilon(v, w)^2 = v^2 w^2 - (v \cdot w)^2,
\end{equation}
where $v^2 = (v^0)^2 - (v^1)^2$ and $v \cdot w = v^0 w^0 - v^1 w^2$.

Therefore,
\begin{equation}
    \label{eq:bubble-gram}
    -\epsilon(q_1, q_2)^2 =
    \det \begin{pmatrix}
        q_1^2 & q_1 \cdot q_2 \\
        q_1 \cdot q_2 & q_2^2
    \end{pmatrix} =
    \frac{1}{4} (4 q_1^2 q_2^2 - (p^2 - q_1^2 - q_2^2)^2).
\end{equation}

Let us first compute the stationary points of $q_2^2$ subject to the constraints mentioned above.  Using the Lagrange multiplier $\lambda_1$, we find
\begin{equation}
    \frac{\partial}{\partial q_2} \Bigl(q_2^2 - \lambda_1 ((p - q_2)^2 - q_1^2)\Bigr) = 0,
\end{equation}
which reads
\begin{equation}
    (1 + \lambda_1) q_2 = \lambda_1 p.
\end{equation}
Using momentum conservation this implies that $q_1 = \frac{1}{1 + \lambda_1} p$, $q_2 = \frac{\lambda_1}{1 + \lambda_1} p$.  We can then determine $\lambda_1$ since $(1 + \lambda_1)^2 = \frac{p^2}{q_1^2}$ so $\lambda_1 = -1 \pm \sqrt{\frac{p^2}{q_1^2}}$.  Using this value of $\lambda_1$ we find
\begin{equation}
    \label{eq:bubble-q2square}
    q_2^2 = (\norm{q_1} \pm \norm{p})^2,
\end{equation}
for the stationary points.

At this stage we don't know yet if these are true minima or maxima.  To decide the nature of the stationary points and find the minima and maxima one can follow the general procedure described in sec.~\ref{sec:two-loop_sunrise} which involves computing a bordered Hessian (see eq.~\eqref{eq:bordered-hessian}).  Their nature depends on the signs of $p^2$ and $q_1^2$.

In this case we do not need the full power of a general theorem since a direct analysis of the inequalities suffices.  Going back to the equation eq.~\eqref{eq:bubble-gram}, we see that for real Lorentz kinematics we have
\begin{equation}
    (q_2^2 - q_1^2 - p^2)^2 \geq 4 q_1^2 p^2.
\end{equation}
Let us assume $p^2 > 0$.  Then, if $q_1^2 < 0$ the inequality is satisfied for all values of $q_2^2$.  The same holds if $p^2 < 0$ and $q_1^2 > 0$.  But if $p^2 > 0$ and $q_1^2 > 0$, then we have
\begin{equation}
    \Bigl(q_2^2 - (\norm{q_1} - \norm{p})^2\Bigr) \Bigl(q_2^2 - (\norm{q_1} + \norm{p})^2\Bigr) > 0.
\end{equation}
It follows that the possible values for $q_2^2$ are either
\begin{equation}
    q_2^2 \geq (\norm{q_1} \pm \norm{p})^2,
\end{equation}
or
\begin{equation}
    q_2^2 \leq (\norm{q_1} \pm \norm{p})^2.
\end{equation}

If instead $p^2 < 0$ and $q_1^2 < 0$, then we have
\begin{equation}
    \Bigl(q_2^2 + (\sqrt{-q_1^2} - \sqrt{-p^2})^2\Bigr) \Bigl(q_2^2 + (\sqrt{-q_1^2} + \sqrt{-p^2})^2\Bigr) > 0.
\end{equation}
Then, either
\begin{equation}
    q_2^2 \geq -(\sqrt{-q_1^2} \pm \sqrt{-p^2})^2,
\end{equation}
or
\begin{equation}
    q_2^2 \leq -(\sqrt{-q_1^2} \pm \sqrt{-p^2})^2.
\end{equation}

Assuming that $p^2 > 0$, for $q_1^2 < 0$ the range of $q_2^2$ is $(-\infty, \infty)$ while for $q_1^2 > 0$ the range of $q_2^2$ is $(-\infty, a_2) \cup (b_2, \infty)$ with $a_2 = (\norm{q_1} - \norm{p})^2$ and $b_2 = (\norm{q_1} + \norm{p})^2$.  The end-points of the integration domain (except the ones at infinity) are the same as obtained by the stationary point study.

The $q_2^2$ integral has two forms
\begin{equation}
    \int_{-\infty}^\infty \frac{d x_2}{(x_2 - c_2) \sqrt{(x_2 - a_2)(x_2 - b_2)}},
\end{equation}
where the roots $a_2$ and $b_2$ are not real, and
\begin{equation}
   \int_{(-\infty, a_2) \cup (b_2, \infty)} \frac{d x_2}{(x_2 - c_2) \sqrt{(x_2 - a_2)(x_2 - b_2)}},
\end{equation}
where the roots $a_2$ and $b_2$ are real with $a_2 < b_2$.

These integrals can be seen as integrals along contours in the curve $y_2^2 = (x_2 - a_2) (x_2 - b_2)$, which is a double cover of the complex $x_2$ plane, branched at two points $x_2 = a_2$ and $x_2 = b_2$.

This curve can be rewritten as $y_2^2 = x_2^2 + \alpha_2 x_2 + \beta_2$ (with $\alpha_2 = -a_2 - b_2$ and $\beta_2 = a_2 b_2$) and after completing the square as $(x_2 + \frac{\alpha_2}{2} - y_2)(x_2 + \frac{\alpha_2}{2} + y_2) = \frac{\alpha_2^2}{4} - \beta_2$.  If we denote $t_2 = x_2 - y_2 + \frac{\alpha_2}{2}$ we have $x_2 + y_2 + \frac{\alpha_2}{2} = \frac{\frac{\alpha_2^2}{4} - \beta_2}{t_2}$ and
\begin{gather}
    x_2 = \frac{1}{2} \Bigl(-\alpha_2 + t_2 + \frac{\frac{\alpha_2^2}{4} - \beta_2}{t_2}\Bigr), \\
    y_2 = \frac{1}{2} \Bigl(\frac{\frac{\alpha_2^2}{4} - \beta_2}{t_2} - t_2\Bigr).
\end{gather}
This provides a rationalization of the curve and is a useful change of variables in the integral.

In terms of the coordinate $t_2$ we have $\frac{d x_2}{\sqrt{(x_2 - a_2)(x_2 - b_2)}} = \frac{d x_2}{y_2} = -\frac{d t_2}{t_2}$.  Then, we have
\begin{equation}
    \omega_2 = \frac{d x_2}{(x_2 - c_2) y_2} =
    -\frac{2 d t_2}{(t_2 - t_2^+)(t_2 - t_2^-)},
\end{equation}
with
\begin{equation}
    \label{eq:t2pm}
    t_2^{\pm}  = c_2 - \frac{a_2 + b_2}{2} \pm \sqrt{(c_2 - a_2)(c_2 - b_2)}.   
\end{equation}
Then we have
\begin{equation}
    \omega_2 = -\frac{2}{t_2^+ - t_2^-} d \log \frac{t_2 - t_2^+}{t_2 - t_2^-} =
    -\frac{1}{\sqrt{(c_2 - a_2)(c_2 - b_2)}} d \log \frac{t_2 - t_2^+}{t_2 - t_2^-}.
\end{equation}
The square root prefactor is now independent on $t_2$ and can be combined with the outer differential form $\frac{d q_1^2}{q_1^2 - m_1^2}$.  We have $c_2 = m_2^2$, $a_2 = (\norm{q_1} - \norm{p})^2$ and $b_2 = (\norm{q_1} + \norm{p})^2$.  We will also set $x_1 = q_1^2$ for brevity.  Replacing these in the square root we find
\begin{equation}
    \frac{d x_1}{(x_1 - c_1) \sqrt{(x_1 - a_1)(x_1 - b_1)}},
\end{equation}
with $c_1 = m_1^2$, $a_1 = (\norm{p} - m_2)^2$ and $b_1 = (\norm{p} + m_2)^2$.  This can be treated in the same way as before.  Indeed, we can introduce a variable $y_1$ defined by $y_1^2 = (x_1 - a_1)(x_1 - b_1)$ and a uniformizing variable $t_1$.
In the end we get
\begin{equation}
    \frac{1}{\sqrt{(m_1^2 - (\norm{p} - m_2)^2)(m_1^2 - (\norm{p} + m_2)^2)}} \int_{\gamma_1} d \log \frac{t_1 - t_1^+}{t_1 - t_1^-} \int_{\gamma_2} d \log \frac{t_2 - t_2^+}{t_2 - t_2^-}.
\end{equation}
The square root in front can be written as
\begin{equation}
    \sqrt{\Delta} = \sqrt{(p^2)^2 + m_1^4 + m_2^4 - 2 p^2 m_1^2 - 2 p^2 m_2^2 - 2 m_1^2 m_2^2},
\end{equation}
which is the familiar K\"all\'en function.

The integration domain is outside the curve in fig.~\ref{fig:bubble_integration_region}.  More precisely, for $q_1^2 \in (-\infty, 0]$ we have $q_2^2 \in (-\infty, \infty)$ while for $q_1^2 \in [0, \infty)$ we have $q_2^2 \in (-\infty, a_2(q_1^2)] \cup [b_2(q_1^2), \infty)$.  For $q_1^2 \in (-\infty, 0]$ we can do the inner integral and find that, as function in $q_1^2$ it has no singularities in the upper half plane (it has a logarithmic branch cut along the positive real axis and there is also a pole at $q_1^2 = m_1^2 - i \epsilon$).  In particular there is no pole at infinity and the contour in $q_1^2$ along the negative real axis can be rotated clockwise to sit on the positive real axis.  Changing the direction of integration introduces a minus sign and combining with the previous integration along $q_1^2 \in [0, \infty)$ produces
\begin{equation}
  -\int_0^\infty \frac {d x_1}{x_1 - c_1} \int_{a_2(x_1)}^{b_2(x_1)} \frac {d x_2}{x_2 - c_2} \frac 1 {\sqrt{(x_2 - a_2)(x_2 - b_2)}}.
\end{equation}
This has the same form as in Euclidean signature, but this form is a result of a cancellation between different regions.

Here we have two integrals but we expect only a single logarithm.  Therefore, it should be possible to do one of the integrals and get a rational multiple of $2 \pi i$.  Since the integration contour in $x_2$ variable goes between $a_2$ and $b_2$, then in the $t_2$ variable it goes between $t_2 = \tilde{t}_2^- = \frac{a_2 - b_2}{2}$ and $t_2 = \tilde{t}_2^+ = \frac{b_2 - a_2}{2}$.  The points $x_2 = a_2$ and $x_2 = b_2$ have a unique point in the double cover and therefore each corresponds to a unique value of $t_2$.  Then, we have $\int_{\gamma_2} d \log \frac{t_2 - t_2^+}{t_2 - t_2^-} = \log \frac{(\tilde{t}_2^+ - t_2^+) (\tilde{t}_2^- - t_2^-)}{(\tilde{t}_2^+ - t_2^-) (\tilde{t}_2^- - t_2^+)}$.  This cross-ratio is very special, since
\begin{equation}
    \frac{(\tilde{t}_2^+ - t_2^+) (\tilde{t}_2^- - t_2^-)}{(\tilde{t}_2^+ - t_2^-) (\tilde{t}_2^- - t_2^+)} =
    \frac{(a_2 - c_2 - \sqrt{(c_2 - a_2)(c_2 - b_2)})(b_2 - c_2 + \sqrt{(c_2 - a_2)(c_2 - b_2)})}{(a_2 - c_2 + \sqrt{(c_2 - a_2)(c_2 - b_2)})(b_2 - c_2 - \sqrt{(c_2 - a_2)(c_2 - b_2)})} = -1,
\end{equation}
where we have used the expressions in eq.~\eqref{eq:t2pm}.

Then, the integral becomes
\begin{equation}
    \pm \frac{\pi i}{\sqrt{\Delta}} \int_{\gamma_1} d \log \frac{t_1 - t_1^+}{t_1 - t_1^-},
\end{equation}
where the sign depends on the determination of the logarithm.

The final answer for the integral resembles the one of the Euclidean signature.  The biggest difference is that we need to replace $p_E^2 \to -p^2$ into
\begin{equation}
    \Delta_E = (p_E^2)^2 + m_1^4 + m_2^4 + 2 p_E^2 m_1^2 + 2 p_E^2 m_2^2 - 2 m_1^2 m_2^2
\end{equation}
to obtain the square root in Lorentzian signature.  This replacement is the same as the one arising from a Wick rotation.

\subsection{Bubble integral in three dimensions}
\label{sec:bubble-3d}

As an example where the inner ``angular'' integral is not zero-dimensional, consider the case of a bubble integral in three dimensions.  As we will see, this integral yields a \emph{simpler} answer than the bubble integral in two dimensions.

For simplicity we compute this integral in Euclidean signature.  We have
\begin{equation}
    \int \frac{d^3 q_1}{(q_1^2 + m_1^2) (q_2^2 + m_2^2)},
\end{equation}
which can be written as
\begin{equation}
    \int_0^\infty \frac{d q_1^2}{q_1^2 + m_1^2} \int_{(\norm{p} - \norm{q_1})^2}^{(\norm{p} + \norm{q_1})^2} \frac{d q_2^2}{q_2^2 + m_2^2} \int \frac{d^3 q_1}{d q_1^2 \wedge d q_2^2}.
\end{equation}
The inner integral is over the space of triangles with side lengths $\norm{p}$, $\norm{q_1}$ and $\norm{q_2}$ in a three-dimensional space, where the vector $p$ is fixed.

We have
\begin{equation}
    \omega_3 = \frac{d^3 q_1}{d q_1^2 \wedge d q_2^2} = \frac{d^3 q_1}{-4 \frac{(q_1 \cdot d q_1) \wedge (q_2 \cdot d q_1) \wedge (v \cdot d q_1)}{v \cdot d q_1}} =
    \frac{v \cdot d q_1}{-4 \epsilon(q_1, q_2, v)}.
\end{equation}
We can choose any vector for $v$, as long as $\epsilon(q_1, q_2, v) \neq 0$.

The one-form $\omega_3$ can be computed by choosing a special coordinate frame where $p = (p_0, 0, 0)$.  Then if we pick $v = (0, 0, 1)$ we find
\begin{equation}
    \omega_3 = \frac{d (q_1)_2}{-4 ((q_1)_0 (q_2)_1 - (q_1)_1 (q_2)_0)} =
    \frac{d (q_1)_2}{4 (q_1)_1 p_0}
\end{equation}
where we have used $(q_2)_1 = -(q_1)_1$, $(q_1)_0 + (q_2)_0 = p_0$.  Using $p_0 = \norm{p}$ and
\begin{gather}
    (q_1)_1 = \sqrt{q_1^2 - \frac{(q_1 \cdot p)^2}{p^2}} \cos \theta, \qquad
    (q_1)_2 = \sqrt{q_1^2 - \frac{(q_1 \cdot p)^2}{p^2}} \sin \theta,
\end{gather}
we finally have that $\omega_3 = -\frac{1}{4 \norm{p}} d \theta$.  The integral over $\theta$ can be done and we obtain $-\frac{\pi}{2 \norm{p}}$.  We can therefore pull out the prefactor directly without going through intermediate integrations as for the two-dimensional bubble.

We are left with the integral
\begin{equation}
    \int_0^\infty \frac{d q_1^2}{q_1^2 + m_1^2} \int_{(\norm{p} - \norm{q_1})^2}^{(\norm{p} + \norm{q_1})^2} \frac{d q_2^2}{q_2^2 + m_2^2}.
\end{equation}
Notice that this integral is odd under $\norm{p} \to -\norm{p}$.  The prefactor also contains a $\norm{p}$ which is also odd under this transformation so overall the integral is invariant under $\norm{p} \to -\norm{p}$.

The inner integral can be computed straightforwardly, which leaves us with
\begin{equation}
    \int_0^\infty \frac{d q_1^2}{q_1^2 + m_1^2} \log \Bigl(\frac{(\norm{p} + \sqrt{q_1^2})^2 + m_2^2}{(\norm{p} - \sqrt{q_1^2})^2 + m_2^2}\Bigr).
\end{equation}
We rewrite the integrand so that it has a cut along the integration region
\begin{equation}
    \int_0^\infty \frac{d q_1^2}{q_1^2 + m_2^2} \log \Bigl(\frac{(\norm{p} - i \sqrt{-q_1^2})^2 + m_2^2}{(\norm{p} + i \sqrt{-q_1^2})^2 + m_2^2}\Bigr).
\end{equation}
Here the square root has the principal determination $\sqrt{z} = \sqrt{\lvert z\rvert} \exp (\frac i 2 \arg{z})$ with $\arg{z} \in (-\pi, \pi)$.  This means that along the real axis $\sqrt{-q_1^2} = i \sqrt{q_1^2}$ and $\Re \sqrt{z} \geq 0$ for all complex $z$.

We define
\begin{equation}
    \rho(q_1^2) = \log \Bigl(\frac{(\norm{p} - i \sqrt{-q_1^2})^2 + m_2^2}{(\norm{p} + i \sqrt{-q_1^2})^2 + m_2^2}\Bigr).
\end{equation}
In principle we could have chosen the function $\rho$ in several other ways, but this form also has the property that $\rho(0) = 0$ and $\lim_{q_1^2 \to \infty} \rho(q_1^2) = 0$.
  
Then the original integral can be written as
\begin{equation}
  \frac 1 2 \int_0^\infty \frac {d q_1^2}{q_1^2 + m_2^2} \operatorname{Disc} \rho(q_1^2),
\end{equation}
where the discontinuity is across the branch cut along the positive real axis and the factor of one half is due to the fact that computing the discontinuity across the branch cut doubles the value of the function $\rho$ just above the cut.

\begin{figure}
  \centering
  \includegraphics{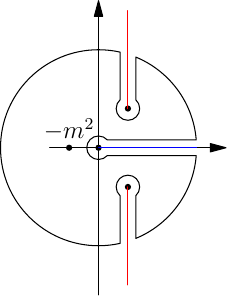}
  \caption{Integration contour for the last integral of the bubble integral in three dimensions.  The horizontal cut is of square root type while the vertical cuts are of logarithmic type.  There is also a pole at $q_1^2 = -m_1^2$.}
  \label{fig:sqrt-cut2}
\end{figure}

We have
\begin{equation}
  \int_0^\infty \frac {d q_1^2}{q_1^2 + m_1^2} \rho(q_1^2) =
  \frac 1 2 \lim_{\epsilon \to 0} \lim_{R \to \infty} \int_{\gamma_1 + \gamma_2 + \gamma_3} \frac {d q_1^2}{q_1^2 + m_1^2} \rho(q_1^2),
\end{equation}
where $\gamma_1$ is a contour from $R$ to $\epsilon$ slightly displaced from the real axis in the lower half plane, $\gamma_2$ is an arc of circle of radius $\epsilon$ and center $0$ which continues the path $\gamma_1$ and $\gamma_3$ is a contour from $\epsilon$ to $R$ slightly displaced in the upper half plane.  This contour can be completed to a closed contour by adding a circle of radius $R$ and center $0$, but paying attention to the logarithmic branch cuts of the function $\rho$.  The resulting contour is sketched in fig.~\ref{fig:sqrt-cut2}.

Let us find the logarithmic branch cuts of $\rho$.  The branch cut condition is
\begin{equation}
  \frac{(\norm{p} - i \sqrt{-q_1^2})^2 + m_2^2}{(\norm{p} + i \sqrt{-q_1^2})^2 + m_2^2} \in \mathbb{R}_-.
\end{equation}
The branch points in $q_1^2$ arise when either the numerator or the denominator of the argument of the logarithm vanish.  This condition can be rewritten as
\begin{equation}
  \Bigl(\bigl(\norm{p} - i \sqrt{-q_1^2}\bigr)^2 + m_2^2\Bigr) \Bigl(\bigl(\norm{p} + i \sqrt{-q_1^2}\bigr)^2 + m_2^2\Bigr) = 0,
\end{equation}
which can be rewritten as
\begin{equation}
  (q_1^2 + m_2^2 - p^2 + 2 i \norm{p} m_2)
  (q_1^2 + m_2^2 - p^2 - 2 i \norm{p} m_2) = 0.
\end{equation}
Solving this we find the logarithmic branch points
\begin{equation}
  q_1^2 = p^2 - m_2^2 \pm 2 i m_2 \norm{p}.
\end{equation}
The equations for the logarithmic branch cuts are more complicated.  We have schematically represented them by the vertical lines going to infinity in fig.~\ref{fig:sqrt-cut2}.  The precise location of the branch cuts is not essential, but we of course require that the integration contours do not cross them.

The integral along the small circle $\gamma_2$ vanishes in the limit $\epsilon \to 0$.  The integral along the pieces of the large circle of radius $R$ also vanishes in the limit $R \to \infty$; the term $\frac {d q_1^2}{q_1^2 + m_1^2}$ contributes a pole at infinity, but its residue vanishes (here we are using the crucial property that $\lim_{q_1^2 \to \infty} \rho(q_1^2) = 0$).

When $q_1^2 = p^2 - m_2^2 + 2 i m_2 \norm{p}$ we have $\sqrt{-q_1^2} = \pm (m_2 - i \norm{p})$ but for the principal determination of the square root we should choose $\sqrt{-q_1^2} = m_2 - i \norm{p}$.  In this case, it is the numerator of the argument of the logarithm in $\rho$ that vanishes.  Since the circles around the logarithmic branch points turn in the clockwise direction, we have that the discontinuity across the branch cut ending at $q_1^2 = p^2 - m_2^2 + 2 i m_2 \norm{p}$ is $-2 \pi i$.  Similarly, the discontinuity across the branch cut ending at $q_1^2 = p^2 - m_2^2 - 2 i m_2 \norm{p}$ is $2 \pi i$.  Therefore, the contribution of the logarithmic branch cuts to the contour integral is
\begin{equation}
  -2 \pi i \int_{p^2 - m_2^2 + 2 i m_2 \norm{p}}^{R_+} \frac {d q_1^2}{q_1^2 + m_1^2} +
  2 \pi i \int_{p^2 - m_2^2 - 2 i m_2 \norm{p}}^{R_-} \frac {d q_1^2}{q_1^2 + m_1^2},
\end{equation}
where $R_\pm$ are some complex numbers whose norm is of order $R$.  They are determined by the intersection of the large circle of radius $R$ and the logarithmic branch cuts.  In the limit $R \to \infty$ this integral becomes
\begin{equation}
  2 \pi i \int_{p^2 - m_2^2 - 2 i m_2 \norm{p}}^{p^2 - m_2^2 + 2 i m_2 \norm{p}} \frac {d q_1^2}{q_1^2 + m_1^2}.
\end{equation}

Equating the value of the integral along the contour in fig.~\ref{fig:sqrt-cut2} with the contribution of the residue at $q_1^2 = -m_1^2$, we obtain
\begin{equation}
  \int_0^\infty \frac {d q_1^2}{q_1^2 + m_1^2} \rho(q_1^2) =
  \pi i \Bigl(
  \log \frac {(\norm{p} - i m_1)^2 + m_2^2}{(\norm{p} + i m_1)^2 + m_2^2} +
  \log \frac {(\norm{p} - i m_2)^2 + m_1^2}{(\norm{p} + i m_2)^2 + m_1^2}
  \Bigr).
\end{equation}
A non-trivial check on the computation is that the answer should be symmetric under the exchange $m_1 \leftrightarrow m_2$.

This can be simplified to
\begin{equation}
  \int_0^\infty \frac {d q_1^2}{q_1^2 + m_1^2} \rho(q_1^2) = 2 \pi i \log \frac{m_1 + m_2 + i \norm{p}}{m_1 + m_2 - i \norm{p}},
\end{equation}
where we assumed $m_1 > 0$, $m_2 > 0$ and $\norm{p} > 0$.

Including the prefactor of $-\frac {\pi}{2 \norm{p}}$ we obtain the final answer for the bubble integral in three dimensions in Euclidean signature
\begin{equation}
  -\frac {\pi^2 i}{\norm{p}} \log \frac{m_1 + m_2 + i \norm{p}}{m_1 + m_2 - i \norm{p}}.
\end{equation}
The result is real and positive, but this is not completely manifest.  It can be rewritten as
\begin{equation}
  \frac {2 \pi^2}{\norm{p}} \arctan \frac {\norm{p}}{m_1 + m_2}.
\end{equation}
This integral is not difficult to compute by other methods (see for example ref.~\cite[eq.~12]{Nickel:1978ds}).  See also ref.~\cite[eq.~3.19]{Hannesdottir:2022bmo} for a recent occurrence of the same integral.

Curiously, this form of the answer is not invariant under $m_i \to -m_i$, which is a symmetry of the original integral.

\section{Triangle integrals}
\label{sec:triangle}

\subsection{Triangle integral in two dimensions}
\label{sec:triangle_2d}

Consider the triangle integral in two dimensions.  This integral is usually computed by reducing it to three bubble integrals.  Each of the bubble integrals has a different prefactor so the integral is not ``pure'' in the language used in the literature.  Such integrals can not be computed by the usual means of taking differentials and they are usually reduced to ``pure'' integrals using a procedure of integral reduction.  In this section, we use this example to illustrate how this new method of direct integration copes with this difficulty.

We want to compute
\begin{equation}
  \int \frac {d^2 q_1}{(q_1^2 + m_1^2) (q_2^2 + m_2^2) (q_3^2 + m_3^2)} =
  \int_0^\infty \frac {d q_1^2}{q_1^2 + m_1^2} \int_{a_2}^{b_2} \frac {d q_2^2}{q_2^2 + m_2^2}
  \frac 1 {q_3^2 + m_3^2} \frac {d^2 q_1}{d q_1^2 \wedge d q_2^2}.
\end{equation}
The integrand of the second integral becomes
\begin{equation}
  \frac 1 {4 \epsilon(q_1, q_2) (q_3^2 + m_3^2)}.
\end{equation}
Given $q_1^2$ and $q_2^2$, the possible values of $q_3^2$ are $(q_3^2)_\pm$, the roots of a quadratic equation arising from the vanishing of the Gram determinant $\det G(q_1, q_2, q_3) = 0$.  Let us write this equation as $\alpha_3 (q_3^2)^2 + \beta_3 q_3^2 + \gamma_3 = 0$, where $\alpha_3$, $\beta_3$ and $\gamma_3$ are polynomials in $p_1^2, p_2^2, p_3^2$ and $q_1^2, q_2^2$.  Then we have
\begin{equation}
  (q_3^2)_\pm = \frac {-\beta_3 \pm \sqrt{\beta_3^2 - 4 \alpha_3 \gamma_3}}{2 \alpha_3}.
\end{equation}

Geometrically, the two configurations of momenta corresponding to the two values of $q_3^2$ where $q_1^2$ and $q_2^2$ are fixed, are related by a reflection so $\epsilon(q_1, q_2)$ changes sign.  We therefore have
\begin{multline}
  \frac 1 {\epsilon(q_1, q_2) (q_3^2 + m_3^2)} =
  \frac 1 {\sqrt{\det G(q_1, q_2)}} \Bigl(\frac 1 {(q_3^2)_+ + m_3^2} - \frac 1 {(q_3^2)_- + m_3^2}\Bigr) =\\=
  - \frac {\sqrt{\beta_3^2 - 4 \alpha_3 \gamma_3}}{\sqrt{\det G(q_1, q_2)}} \frac 1 {\alpha_3 m_3^4 - \beta_3 m_3^2 + \gamma_3}.
\end{multline}

As before, we have $a_2 = \inf q_2^2 = (\norm{p_3} - \norm{q_1})^2$ and $b_2 = \sup q_2^2 = (\norm{p_3} + \norm{q_1})^2$.  When $q_1$ and $q_2$ are collinear $q_2^2$ satisfies the equation $(q_2^2 - a_2) (q_2^2 - b_2) = 0$ which when expanded yields
\begin{equation}
  (p_3^2)^2 + (q_1^2)^2 + (q_2^2)^2 - 2 p_3^2 q_1^2 - 2 p_3^2 q_2^2 - 2 q_1^2 q_2^2 = 0.
\end{equation}
When this equation holds we must have $\beta_3^2 - 4 \alpha_3 \gamma_3 = 0$ since the two configurations for $q_3^2$ arise via a reflection in the line supporting the momentum $p_3$.  But when the momenta $q_1$ and $q_2$ are collinear with $p_3$, the reflection does not produce a new solution.

At the same time, we have
\begin{equation}
  4 \det G(q_1, q_2) = (p_3^2)^2 + (q_1^2)^2 + (q_2^2)^2 - 2 p_3^2 q_1^2 - 2 p_3^2 q_2^2 - 2 q_1^2 q_2^2.
\end{equation}
Hence, when $q_2^2 = a_2$ or $q_2^2 = b_2$ both the numerator $\sqrt{\beta_3^2 - 4 \alpha_3 \gamma_3}$ and the denominator factor $\sqrt{\det G(q_1, q_2)}$ vanish.  An explicit computation reveals that
\begin{multline}
  \beta_3^2 - 4 \alpha_3 \gamma_3 =
  \frac 1 {16} \Bigl((p_1^2)^2 + (p_2^2)^2 + (p_3^2)^2
  -2 p_1^2 p_2^2 - 2 p_1^2 p_3^2 - 2 p_2^2 p_3^2\Bigr) \\
  \Bigl((q_1^2)^2 + (q_2^2)^2 + (p_3^2)^2
  -2 q_1^2 p_3^2 - 2 q_2^2 p_3^2 - 2 q_1^2 q_2^2\Bigr).
\end{multline}
Therefore, we can pull out a factor
\begin{equation}
  -\frac 1 2 \sqrt{(p_1^2)^2 + (p_2^2)^2 + (p_3^2)^2
  -2 p_1^2 p_2^2 - 2 p_1^2 p_3^2 - 2 p_2^2 p_3^2}
\end{equation}
dependent only on external kinematics and we are left with the integrals
\begin{equation}
  \int_0^\infty \frac {d q_1^2}{q_1^2 + m_1^2} \int_{a_2}^{b_2} \frac {d q_2^2}{q_2^2 + m_2^2}
\frac 1 {\alpha_3 m_3^4 - \beta_3 m_3^2 + \gamma_3}.
\end{equation}
The integral in $q_2^2$ can be done by straightforward partial fractioning in $q_2^2$.  If $\alpha_3 m_3^4 - \beta_3 m_3^2 + \gamma_3 = \alpha_2 (q_2^2)^2 + \beta_2 q_2^2 + \gamma_2$ and this polynomial in $q_2^2$ has roots $x_\pm$, then
\begin{multline}
  \label{eq:triangle-2d-second-int}
  \int_{a_2}^{b_2} \frac {d q_2^2}{q_2^2 + m_2^2} \frac 1 {\alpha_3 m_3^4 - \beta_3 m_3^2 + \gamma_3} =
  \int_{a_2}^{b_2} \frac {d q_2^2}{q_2^2 + m_2^2} \frac 1 {\alpha_2 (q_2^2 - x_+)(q_2^2 - x_-)} = \\
  \frac 1 {\alpha_2 m_2^4 - \beta_2 m_2^2 + \gamma_2} \log \frac {b_2 + m_2^2}{a_2 + m_2^2} + \\
  \frac 1 {x_+ + m_2^2} \frac 1 {\sqrt{\beta_2^2 - 4 \alpha_2 \gamma_2}} \log \frac {b_2 - x_+}{a_2 - x_+} - \\
  \frac 1 {x_- + m_2^2} \frac 1 {\sqrt{\beta_2^2 - 4 \alpha_2 \gamma_2}} \log \frac {b_2 - x_-}{a_2 - x_-}.
\end{multline}

An explicit calculation yields
\begin{multline}
  \beta_2^2 - 4 \alpha_2 \gamma_2 =
  \frac 1 {16} \Bigl((p_1^2)^2 + (p_2^2)^2 + (p_3^2)^2
  -2 p_1^2 p_2^2 - 2 p_1^2 p_3^2 - 2 p_2^2 p_3^2\Bigr) \\
  \Bigl(m_2^4 + (p_2^2)^2 + (q_1^2)^2 + 2 m_2^2 p_2^2 + 2 m_2^2 q_1^2 - 2 p_2^2 q_1^2\Bigr).
\end{multline}
We also have that
\begin{equation}
  \frac 1 {\alpha_2 m_2^4 - \beta_2 m_2^2 + \gamma_2} +
  \frac 1 {x_+ + m_2^2} \frac 1 {\sqrt{\beta_2^2 - 4 \alpha_2 \gamma_2}} -
  \frac 1 {x_- + m_2^2} \frac 1 {\sqrt{\beta_2^2 - 4 \alpha_2 \gamma_2}} = 0.
\end{equation}

To do the final integral, we proceed as before.  First, notice that when $\norm{q_1} = 0$ or $\norm{q_1} \to \infty$ we have that $a_2$ and $b_2$ coincide.  This means that the logarithms vanish.  In fact, the integrand of the $q_1^2$ integral has a square root branch point at $q_1^2 = 0$ and $q_1^2 \to \infty$.  It looks like the result of the integration in eq.~\eqref{eq:triangle-2d-second-int} (and therefore the integrand for the $q_1^2$ integration) also has square root branch points at $\beta_2^2 - 4 \alpha_2 \gamma_2 = 0$.  Interestingly, these square root branch points are actually canceled in the combination in eq.~\eqref{eq:triangle-2d-second-int}.  Indeed, when taking $q_1^2$ along a path such that $\beta_2^2 - 4 \alpha_2 \gamma_2$ goes once around the origin, we have $x_{+} \leftrightarrow x_{-}$ and $\sqrt{\beta_2^2 - 4 \alpha_2 \gamma_2} \to -\sqrt{\beta_2^2 - 4 \alpha_2 \gamma_2}$.  Then the expression in eq.~\eqref{eq:triangle-2d-second-int} is sent to itself.

As before, we have $\alpha_2 m_2^4 - \beta_2 m_2^2 + \gamma_2 = \alpha_1 (q_1^2)^2 + \beta_1 q_1^2 + \gamma_1$.  This time $\alpha_1$, $\beta_1$ and $\gamma_1$ depend only on the external momenta and the masses.

The integrand has some pole singularities.  Indeed, $\alpha_2 m_2^4 - \beta_2 m_2^2 + \gamma_2 = \alpha_1 (q_1^2)^2 + \beta_1 q_1^2 + \gamma_1 = 0$ when $q_1^2 = (q_1^2)_\pm$.  We have
\begin{equation}
    \operatorname{Res}_{q_1^2 = (q_1^2)_{\pm}} \frac{1}{\alpha_1 (q_1^2)^2 + \beta_1 q_1^2 + \gamma_1} =
    \operatorname{Res}_{q_1^2 = (q_1^2)_{\pm}} \frac{1}{\alpha_1 (q_1^2 - (q_1^2)_{+})(q_1^2 - (q_1^2)_{-})} =
    \frac{\pm 1}{\sqrt{\beta_1^2 - 4 \alpha_1 \gamma_1}}.
\end{equation}
There is also a pole when $x_{+} + m_2^2 = 0$.  Since $\alpha_2 (x_+ + m_2^2)(x_- + m_2^2) = \alpha_1 (q_1^2 - (q_1^2)_+)(q_1^2 - (q_1^2)_-)$ when $x_{+} + m_2^2 = 0$ we have that either $q_1^2 = (q_1^2)_{+}$ or $q_1^2 = (q_1^2)_{-}$.  Let us assume for definiteness that we have $q_1^2 = (q_1^2)_{+}$.  Then,
\begin{multline}
    \operatorname{Res}_{q_1^2 = (q_1^2)_+} \frac{1}{x_+ + m_2^2} =
    \lim_{q_1^2 \to (q_1^2)_+} \frac{(q_1^2 - (q_1^2)_+)(q_1^2 - (q_1)_-)}{(x_+ + m_2^2)(x_- + m_2^2)} \frac{x_- + m_2^2}{q_1^2 - (q_1^2)_{-}} = \\ =
    \frac{\alpha_2}{\alpha_1} \frac{x_{-} - x_{+}}{(q_1^2)_{+} - (q_1^2)_{-}} =
    -\frac{\sqrt{\beta_2^2 - 4 \alpha_2 \gamma_2}}{\sqrt{\beta_1^2 - 4 \alpha_1 \gamma_1}},
\end{multline}
where we have used $m_2^2 = -x_{+}$.

The final integral to do is over $q_1^2$ and runs from $0$ to $\infty$.  This along a square root branch cut so the integral can be written as one half the integral around the branch cut.  This contour can be deformed to a sum of contours around the poles described above and around the logarithmic branch cuts.  The integrals around the logarithmic branch cuts can be written as integrals along a contour connecting the two logarithmic branch points where the new integrand is obtained by replacing the logarithm by $2 \pi i$.  After this replacement the integral becomes easy to perform, by using identities such as eq.~\eqref{eq:known-integral}.

We will not go through all the trivial (but tedious) steps in detail (see sec.~\ref{sec:bubble-3d} for more details on the method), except to describe the determination of the locations of logarithmic branch points.  From eq.~\eqref{eq:triangle-2d-second-int} we have that the first logarithmic branch points appear for values of $q_1^2$ where $a_2 = -m_2^2$ and $b_2 = -m_2^2$.  Using $a_2 = (\norm{p_2} - \norm{q_1})^2$ and $b_2 = (\norm{p_2} + \norm{q_1})^2$ we find logarithmic branch points at $q_1^2 = (\norm{p_3} \pm i m_2)^2$.  From the second and third terms we have $x_\pm = b_2$ which implies that $\alpha_2 b_2^2 + \beta_2 b_2 + \gamma_2 = 0$.  When written as an equation in $\norm{q_1}$ this equation is of degree four, so one might worry that we need to deal with roots of order four.  Fortunately, it turns out that this degree four equation is very special and its roots can be written using a single square root
\begin{equation}
  \norm{q_1} = \frac {p_1^2 - p_2^2 - p_3^2 \pm \sqrt{D}}{2 \norm{p_3}},
\end{equation}
where
\begin{equation}
  D = (p_1^2)^2 + (p_2^2)^2 + (p_3^2)^2 - 2 p_1^2 p_2^2 - 2 p_1^2 p_3^2 - 2 p_2^2 p_3^2 - 4 m_3^2 p_3^2.
\end{equation}
Each one of these roots appears with multiplicity two.  A similar analysis applies for $a_2$.

In this section we have demonstrated an algorithm for computing a reducible Feynman integral with multiple prefactors, without performing an integral reduction.  Integral reductions (see ref.~\cite{Laporta:2004rb}) are often resource-intensive and require a non-canonical and often symmetry-breaking choice of basis.

One interesting fact we can notice is the occurrence of higher order equations, but so far such that their roots only require quadratic field extensions.  It is plausible that by this method of integration no unnecessary field extensions will be required.  When computing a particular integral in ref.~\cite{Bourjaily:2019igt} using \texttt{HyperInt} (see ref.~\cite{Panzer:2014caa}) we encountered field extensions of degree $16$ while the final answer was completely rational.  Higher order equations are expected to occur (see refs.~\cite{Mizera:2021icv, Bourjaily:2022vti}) in general for more complicated integrals.

\subsection{Triangle integral in three dimensions}

\begin{figure}
    \centering
    \includegraphics{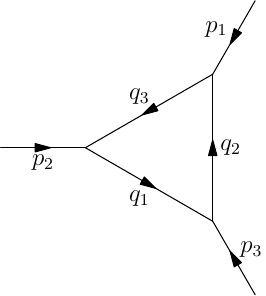}
    \caption{Kinematics of the triangle integral.}
    \label{fig:triangle_integral}
\end{figure}

It is instructive to consider the triangle integral in three dimensions.\footnote{Often the integrals are studied in Feynman parameter space, not in momentum space as we do here.  In Feynman parameter space the integrals in odd dimensions may contain square roots, which complicates their analysis.  One of the advantages of the momentum space approach is that it can be done in even or odd dimensions with no differences.}  In this case we expect the answer to contain one logarithm, so we should be able to compute \emph{two} integrals as rational multiples of $2 \pi i$.

For the triangle integral we have three external momenta $p_1, p_2, p_3$ with $p_1 + p_2 + p_3 = 0$ and three internal momenta $q_1, q_2, q_3$ such that $p_1 = q_3 - q_2$, $p_2 = q_1 - q_3$ and $p_3 = q_2 - q_1$ (see fig.~\ref{fig:triangle_integral}).

We do the integral in Euclidean signature.  The integral reads
\begin{equation}
    \label{eq:triangle-integral}
    \int \frac{d^3 q_1}{(q_1^2 + m_1^2) (q_2^2 + m_2^2) (q_3^2 + m_3^2)}.
\end{equation}
Putting the integral in the Cutkosky form we find
\begin{equation}
    \int_{a_1}^{b_1} \frac{d q_1^2}{q_1^2 + m_1^2} \int_{a_2}^{b_2} \frac{d q_2^2}{q_2^2 + m_2^2} \int_{a_3}^{b_3} \frac{d q_3^2}{q_3^2 + m_3^2} \Bigl(\frac{d^3 q_1}{d q_1^2 \wedge d q_2^2 \wedge d q_3^2}\Bigr).
\end{equation}

We will determine $a_1, a_2, a_3$ and $b_1, b_2, b_3$ in the following, but first we compute
\begin{equation}
    \frac{d^3 q_1}{d q_1^2 \wedge d q_2^2 \wedge d q_3^2}.
\end{equation}
Using momentum conservation we find $q_2 = q_1 + p_3$ and $q_3 = q_1 + p_1 + p_3$ which implies that $d q_2^2 = 2 q_2 \cdot d q_1$ and $d q_3^2 = 2 q_3 \cdot d q_1$ since the external momenta are taken to be constant.

Next, we find $d q_1^2 \wedge d q_2^2 \wedge d q_3^2 = 8 \epsilon(q_1, q_2, q_3) d^3 q_1$.  We have
\begin{equation}
    \epsilon(q_1, q_2, q_3)^2 =
    \det \begin{pmatrix}
        q_1^2 & q_1 \cdot q_2 & q_1 \cdot q_3 \\
        q_1 \cdot q_2 & q_2^2 & q_2 \cdot q_3 \\
        q_1 \cdot q_3 & q_2 \cdot q_3 & q_3^2
    \end{pmatrix}.
\end{equation}
Therefore, we have
\begin{equation}
  \frac{d^3 q_1}{d q_1^2 \wedge d q_2^2 \wedge d q_3^2} = \frac 1{8 \epsilon(q_1, q_2, q_3)} = \frac 1 {8 \sqrt{\det (q_i \cdot q_j)_{1 \leq i, j \leq 3}}}.
\end{equation}

To compute the boundary values, $a_i$ and $b_i$ we proceed as follows.  First, we have $a_1 = 0$ and $b_1 = \infty$.  Next, have the same problem as for the bubble case in sec.~\ref{sec:bubble}.  That is, given the fixed value of $q_1^2$ and $p_3 = q_2 - q_1$, find the extremal values of $q_2^2$.  Finally, for the extremal values of $q_3^2$ there are several possibilities.  First, there is a constraint arising from $q_3^2 = (q_1 - p_2)^2$, or from the bubble with momenta $q_1, q_3$.  Second, there is a constraint arising from the bubble with momenta $q_2, q_3$.  Taken together, these imply the triangle Landau equations.

Thus, we see the hierarchical principle of ref.~\cite{landshoff_hierarchical_1966} arise in a quite concrete way.  What is not so clear are singularities in the bubble with momenta $q_2, q_3$, for example.

Let us denote $\alpha_3 (q_3^2)^2 + \beta_3 q_3^2 + \gamma_3 = \det (q_i \cdot q_j)_{1 \leq i, j \leq 3}$, where $\alpha_3$, $\beta_3$ and $\gamma_3$ are complicated polynomials which we will not need to spell out.  We define $c_3 = \frac{\beta_3}{\alpha_3}$, $d_3 = \frac{\gamma_3}{\alpha_3}$ and $x_3 = q_3^2$, $y_3^2 = x_3^2 + c_3 x_3 + d_3$.  This curve can be parametrized by
\begin{gather}
    t_3 = x_3 + \frac{c_3}{2} - y_3, \qquad
    \frac{\Delta_3}{4 t_3} = x_3 + \frac{c_3}{2} + y_3,
\end{gather}
where $\Delta_3 = c_3^2 - 4 d_3$.  This implies that
\begin{equation}
    \omega_3 = \frac{d q_3^2}{(q_3^2 + m_3^2) \sqrt{\alpha_3 (q_3^2)^2 + \beta_3 q_3^2 + \gamma_3}}
    = -\frac{1}{\sqrt{\alpha_3 (m_3^2)^2 - \beta_3 m_3^2 + \gamma_3}} d \log \frac{t_3 - t_3^+}{t_3 - t_3^-},
\end{equation}
for
\begin{equation}
    t_3^\pm = \frac{c_3}{2} - m_3^2 \pm \sqrt{(m_3^2)^2 - c_3 m_3^2 + d_3}.
\end{equation}
The boundaries of the integral in $x_3$ are for values of $t_3$ where $y_3$ vanishes.  This means that $t_3^2 = \frac{\Delta_3}{4}$ therefore $t_{3, i} = -\frac 1 2 \sqrt{\Delta_3}$ and $t_{3, f} = \frac 1 2 \sqrt{\Delta_3}$.  Computing the definite integral involves computing a cross-ratio
\begin{equation}
  \frac{(t_{3, f} - t_3^+) (t_{3, i} - t_3^-)}{(t_{3, f} - t_3^-) (t_{3, i} - t_3^+)} = -1,
\end{equation}
which follows from
\begin{equation}
  (t_{3, f} - t_3^+) (t_{3, i} - t_3^-) =
  -\frac{\Delta_3}4 - \frac{\sqrt{\Delta_3}}2 (t_3^- - t_3^+) + \frac{\Delta_3}4,
\end{equation}
and similarly for the denominator.  This integral therefore will produce a constant transcendental factor of $\pi$.  The final answer is
\begin{equation}
  \int_{a_3}^{b_3} \frac{d q_3^2}{(q_3^2 + m_3^2) \sqrt{\alpha_3 (q_3^2)^2 + \beta_3 q_3^2 + \gamma_3}} =
  \frac {\pi}{\sqrt{-\alpha_3 m_3^4 + \beta_3 m_3^2 - \gamma_3}}.
\end{equation}
This integral is positive if the quantity under the square root in the integrand is positive for $q_3^2 \in (a_3, b_3)$ and $-m_3^2 \not\in (a_3, b_3)$ (this is certainly the case if $m_3^2 > 0$ since $a_3, b_3 \geq 0$).  Then the quantity under the square root in the answer is positive.

Now we want to do the second integral,
\begin{equation}
  \int_{a_2}^{b_2} \frac {d q_2^2}{q_2^2 + m_2^2} \frac {\pi}{\sqrt{-\alpha_3 m_3^4 + \beta_3 m_3^2 - \gamma_3}}.
\end{equation}
The quantity under the square root is minus the Gram determinant of $q_1, q_2, q_3$, evaluated at the ``Euclidean on-shell'' condition $q_3^2 = -m_3^2$.  We now have $-\alpha_3 m_3^4 + \beta_3 m_3^2 - \gamma_3 = \alpha_2 (q_2^2)^2 + \beta_2 q_2^2 + \gamma_2$.

We have
\begin{equation}
  \alpha_2 (q_2^2)^2 + \beta_2 q_2^2 + \gamma_2 =
  \alpha_2 \Bigl(q_2^2 + \frac {\beta_2}{2 \alpha_2}\Bigr)^2 -
  \frac {\Delta_2}{4 \alpha_2},
\end{equation}
where
\begin{multline}
  \Delta_2 = \beta^2 - 4 \alpha_2 \gamma_2 =
  \frac 1 {16} (m_3^4 + 2 m_3^2 p_2^2 + 2 m_3^2 q_1^2 + (p_2^2)^2 - 2 p_2^2 q_1^2 + (q_1^2)^2) \\
  ((p_1^2)^2 - 2 p_1^2 p_2^2 + (p_2^2)^2 - 2 p_1^2 p_3^2 - 2 p_2^2 p_3^2 + (p_3^2)^2).
\end{multline}
The first term can be rewritten as
\begin{equation}
  (q_1^2 - p_2^2 + m_3^2)^2 + 4 m_3^2 p_2^2 > 0.
\end{equation}
The second term, up to a prefactor, is the Gram determinant of $p_1$ and $p_2$.  Indeed,
\begin{equation}
  \det
  \begin{pmatrix}
    p_1^2 & p_1 \cdot p_2 \\
    p_1 \cdot p_2 & p_2^2
  \end{pmatrix}
  = -\frac 1 4 ((p_1^2)^2 - 2 p_1^2 p_2^2 + (p_2^2)^2 - 2 p_1^2 p_3^2 - 2 p_2^2 p_3^2 + (p_3^2)^2).
\end{equation}
This Gram determinant, representing a volume in Euclidean space, is positive hence the second term in the factorization of $\Delta_2$ is negative.  Therefore, $\Delta_2 < 0$ and the roots of $\alpha_2 (q_2^2)^2 + \beta_2 q_2^2 + \gamma_2 = 0$ are complex.  In particular, $\alpha_2 (q_2^2)^2 + \beta_2 q_2^2 + \gamma_2 > 0$ in the region of integration.

If $P$ is a quadratic polynomial, the integral can be rewritten as (see eq.~\eqref{eq:known-integral})
\begin{multline}
    \int_a^b \frac{d x}{x + c} \frac{1}{\sqrt{P(x)}} =
    \frac{1}{\sqrt{P(-c)}} \Bigl\lbrace
    - \log (a + c - \sqrt{p(a)} + \sqrt{p(-c)}) \\
    + \log (-a - c + \sqrt{p(a)} + \sqrt{p(-c)})
    + \log (b + c - \sqrt{p(b)} + \sqrt{p(-c)}) \\
    - \log (-b - c + \sqrt{p(b)} + \sqrt{p(-c)})
    \Bigr\rbrace,
\end{multline}
with $p(x) = P(x) / \text{lc}(P)$ and $\text{lc}(P)$ is the leading coefficient of the polynomial $P$.

Using this formula we can do the $q_2^2$ integral, by substituting $a \to a_2$, $b \to b_2$, $x \to q_2^2$, $c \to m_2^2$, $p \to P_2$ with $P_2(x) = \alpha_2 x^2 + \beta_2 x + \gamma_2$ and $p_2(x) = P_2(x)/\alpha_2$.  The $q_2^2$ integral yields an expression which contains $\sqrt{p_2(a_2)}$ and $\sqrt{p_2(b_2)}$ where $P_2(x) = \alpha_2 x^2 + \beta_2 x + \gamma_2$, $a_2 = (\norm{q_1} - \norm{p_3})^2$ and $b_2 = (\norm{q_1} + \norm{p_3})^2$.  These square roots may produce square branch points at the solutions of $p_2(b_2) = 0$ and $p_2(a_2) = 0$ in the variable $\norm{q_1}$.  These are fourth order equations in $\norm{q_1}$.  However, the polynomials $p_2(b_2)$ and $p_2(a_2)$ are very special and in fact they are perfect squares!  Indeed, we have
\begin{multline}
  P_2(a_2) = 
    \frac{1}{4} \Bigl(m_3^2 \norm{p_3} + p_2^2 \norm{p_3} + (p_1^2 - p_2^2 - p_3^2) \norm{q_1} + \norm{p_3} q_1^2\Bigr)^2 = \frac 1 {64 p_3^2} \times \\
    \left(\sqrt{-4 m_3^{2} p_3^2 +(p_1^2)^{2}-2 p_1^2 p_2^2 -2 p_1^2 p_3^2 + (p_2^2)^{2}-2 p_2^2 p_3^2 + (p_3^2)^{2}} - p_1^2 + p_2^2 + p_3^2 - 2 \norm{p_3} \norm{q_1}\right)^{2} \\ \left(\sqrt{-4 m_3^{2} p_3^2 + (p_1^2)^{2}-2 p_1^2 p_2^2 -2 p_1^2 p_3^2 + (p_2^2)^{2}-2 p_2^2 p_3^2 + (p_3^2)^{2}} + p_1^2 - p_2^2 - p_3^2 + 2 \norm{p_3} \norm{q_1} \right)^{2}.
\end{multline}
\begin{multline}
  P_2(b_2) =
      \frac{1}{4} \Bigl(m_3^2 \norm{p_3} + p_2^2 \norm{p_3} - (p_1^2 - p_2^2 - p_3^2) \norm{q_1} + \norm{p_3} q_1^2\Bigr)^2  =
    \frac 1 {64 p_3^2} \times \\ \left(\sqrt{-4 m_3^{2} p_3^2 +(p_1^2)^{2}-2 p_1^2 p_2^2 -2 p_1^2 p_3^2 + (p_2^2)^{2}-2 p_2^2 p_3^2 + (p_3^2)^{2}} - p_1^2 + p_2^2 + p_3^2 + 2 \norm{p_3} \norm{q_1}\right)^{2} \\ \left(\sqrt{-4 m_3^{2} p_3^2 + (p_1^2)^{2}-2 p_1^2 p_2^2 -2 p_1^2 p_3^2 + (p_2^2)^{2}-2 p_2^2 p_3^2 + (p_3^2)^{2}} + p_1^2 - p_2^2 - p_3^2 - 2 \norm{p_3} \norm{q_1} \right)^{2}.
\end{multline}
It follows that the branch points due to $\sqrt{p_2(a_2)}$ and $\sqrt{p_2(b_2)}$ are in fact not there at all and the square roots can be extracted.

The final integral over $q_1^2$ can be written as an integral along the branch cut in $\sqrt{-q_1^2}$ going from zero to infinity along the positive real axis.  As described above, there are several contributions: the contributions from integrating along the logarithmic branch cuts and the contribution from the pole at $q_1^2 = -m_1^2$.

The logarithmic branch points are at
\begin{gather}
    b_2 + m_3^2 = 0, \qquad \norm{q_1} = - \norm{p_3} \pm i m_3, \\
    a_2 + m_3^2 = 0, \qquad \norm{q_1} = \norm{p_3} \pm i m_3, \\
    \beta_2^2 - 4 \alpha_2 \gamma_2 = 0, \qquad \norm{q_1} = \pm (m_3 + i \norm{p_2}), \quad \norm{q_1} = \pm (m_3 - i \norm{p_2}).
\end{gather}
Not all these branch points are on the main sheet we restricted to.  Indeed, $\norm{q_1} = \sqrt{q_1^2}$ and on the main sheet $\Re \sqrt{z} > 0$, so we should pick the solutions $\norm{q_1} = \norm{p_3} \pm i m_3$ and $\norm{q_1} = m_3 \pm i \norm{p_2}$ and these become the boundaries of integrals along the logarithmic branch cut.  The rest of the calculation is routine and we will not reproduce it.

In ref.~\cite{Nickel:1978ds} this integral is computed by a clever application of partial fractioning followed by contour integration.  The point of the method described here is that no clever partial fractioning is required but instead one can follow a simple algorithm.

\section{The two-loop sunrise integral}
\label{sec:two-loop_sunrise}

\begin{figure}
    \centering
    \includegraphics{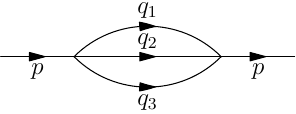}
    \caption{The kinematics of the two-loop sunrise integral.}
    \label{fig:two-loop_sunrise_integral}
\end{figure}

This integral is famously elliptic, see refs.~\cite{Laporta:2004rb, Muller-Stach:2011qkg, Bloch:2013tra}.  We will see the elliptic curve and the holomorphic differential appear explicitly.

We will do the integrals in Euclidean signature.  We start with the integral (see fig.~\ref{fig:two-loop_sunrise_integral})
\begin{equation}
    I = \int \frac{d^2 q_1 d^2 q_2}{(q_1^2 + m_1^2) (q_2^2 + m_2^2) (q_3^2 + m_3^2)}.
\end{equation}
Using Cutkosky's change of variables we have
\begin{equation}
    I = \int_0^\infty \frac{d q_1^2}{q_1^2 + m_1^2} \int_0^\infty \frac{d q_2^2}{q_2^2 + m_2^2} \int_{(q_3^2)_{\text{min}}}^{(q_3^2)_{\text{max}}} \frac{d q_3^2}{q_3^2 + m_3^2} \int_{\gamma} \frac{d^2 q_1 d^2 q_2}{d q_1^2 \wedge d q_2^2 \wedge d q_3^2},
\end{equation}
where $p$ is the external momentum and $p = q_1 + q_2 + q_3$, where $q_1$, $q_2$ and $q_3$ are the momenta through the internal lines of the sunrise diagram.  The values $(q_3^2)_{\text{min}}$ and $(q_3^2)_{\text{max}}$ are obtained by computing the stationary points of $q_3^2$ subject to the constraints that $p = q_1 + q_2 + q_3$ and $q_1^2$ and $q_2^2$ take fixed values.  The contour $\gamma$ is a real cycle on an elliptic curve which we will describe in more detail in the following.

Indeed, it is clear that $q_1^2$ and $q_2^2$ can take any values, but the values of $q_3^2$ are constrained.  For clarity, let us denote the fixed values for $q_1^2$ and $q_2^2$ by $q_1^2 = z_1$ and $q_2^2 = z_2$.  Then, we look for the stationary points of $q_3^2 = (p - q_1 - q_2)^2$ subject to the constraints $q_1^2 = z_1$ and $q_2^2 = z_2$ where $z_1$, $z_2$ are some fixed values.

We define a function
\begin{equation}
    F(q_1, q_2, \lambda_1, \lambda_2) = (p - q_1 - q_2)^2 - \lambda_1 (q_1^2 - z_1) - \lambda_2 (q_2^2 - z_2),
\end{equation}
where $\lambda_1$ and $\lambda_2$ are Lagrange multipliers.
The stationary point conditions read
\begin{gather}
    \frac{\partial}{\partial q_1} \Bigl((p - q_1 - q_2)^2 - \lambda_1 (q_1^2 - z_1) - \lambda_2 (q_2^2 - z_2)\Bigr) = 0, \\
    \frac{\partial}{\partial q_2} \Bigl((p - q_1 - q_2)^2 - \lambda_1 (q_1^2 - z_1) - \lambda_2 (q_2^2 - z_2)\Bigr) = 0,
\end{gather}
while the derivatives with respect to the Lagrange multipliers reproduce the constraints.

Taking the derivatives we find
\begin{gather}
    p - q_1 - q_2 + \lambda_1 q_1 = 0, \\
    p - q_1 - q_2 + \lambda_2 q_2 = 0.
\end{gather}
In particular, this implies that $\lambda_1 q_1 = \lambda_2 q_2$.  By squaring we have $\lambda_1^2 z_1 = \lambda_2^2 z_2$ so $\lambda_2 = \pm \lambda_1 \sqrt{\frac{z_1}{z_2}}$.

Let us first assume that $\lambda_1$, $\lambda_2$ are non-vanishing.  The system of equations above has a solution
\begin{gather}
    q_1 = \frac{\lambda_2}{\lambda_1 + \lambda_2 - \lambda_1 \lambda_2} p, \qquad
    q_2 = \frac{\lambda_1}{\lambda_1 + \lambda_2 - \lambda_1 \lambda_2} p.
\end{gather}
By squaring the first equation we find
\begin{equation}
    \Bigl(1 + \frac{\lambda_2}{\lambda_1} - \lambda_2\Bigr)^2 = \frac{p^2}{z_2}.
\end{equation}
Using $\frac{\lambda_2}{\lambda_1} = \pm_1 \sqrt{\frac{z_1}{z_2}}$ we find
\begin{gather}
    \lambda_1 = 1 \pm_1 \sqrt{\frac{z_2}{z_1}} \pm_3 \sqrt{\frac{p^2}{z_1}}, \qquad
    \lambda_2 = 1 \pm_1 \sqrt{\frac{z_1}{z_2}} \pm_2 \sqrt{\frac{p^2}{z_2}},
\end{gather}
with $(\pm_1) (\pm_2) (\pm_3) = 1$.  Then, we have
\begin{gather}
    q_1 = \mp_3 \sqrt{\frac{z_1}{p^2}} p, \qquad
    q_2 = \mp_2 \sqrt{\frac{z_2}{p^2}} p, \\
    q_3 = \Bigl(1 \pm_3 \sqrt{\frac{z_1}{p^2}} \pm_2 \sqrt{\frac{z_2}{p^2}}\Bigr) p,
\end{gather}
hence
\begin{equation}
    q_3^2 = (\norm{p} \pm_3 \sqrt{z_1} \pm_2 \sqrt{z_2})^2.
\end{equation}

In conclusion, we find four different critical points for $q_3^2$.  Let us now study their nature in more detail.  In order to decide the nature of the critical points, we compute the bordered Hessian matrix
\begin{equation}
  \label{eq:bordered-hessian}
    H(F) =
    \begin{pmatrix}
        0 & 0 & \frac{\partial^2 F}{\partial \lambda_1 \partial q_1} & \frac{\partial^2 F}{\partial \lambda_1 \partial q_2} \\
        0 & 0 & \frac{\partial^2 F}{\partial \lambda_2 \partial q_1} & \frac{\partial^2 F}{\partial \lambda_2 \partial q_2} \\
        \frac{\partial^2 F}{\partial q_1 \partial \lambda_1} & \frac{\partial^2 F}{\partial q_1 \partial \lambda_2} & \frac{\partial^2 F}{\partial q_1 \partial q_1} & \frac{\partial^2 F}{\partial q_1 \partial q_2} \\
        \frac{\partial^2 F}{\partial q_2 \partial \lambda_1} & \frac{\partial^2 F}{\partial q_2 \partial \lambda_2} & \frac{\partial^2 F}{\partial q_2 \partial q_1} & \frac{\partial^2 F}{\partial q_2 \partial q_2}
    \end{pmatrix}.
\end{equation}
In our example we have
\begin{equation}
    H(F) =
    \begin{pmatrix}
        0 & 0 & -2 q_1 & 0 \\
        0 & 0 & 0 & -2 q_2 \\
        -2 q_1 & 0 & 2 (1 - \lambda_1) \;\mathbf{1}_2 & 2 \;\mathbf{1}_2 \\
        0 & -2 q_2 & 2 \;\mathbf{1}_2 & 2 (1 - \lambda_2) \;\mathbf{1}_2
    \end{pmatrix}.
\end{equation}

We have four variables (the components of $q_1$ and $q_2$) and two constraints.  Therefore, we need to look at the signs of the $5 \times 5$ and $6 \times 6$ principal minors.  Computing these minors we find
\begin{gather}
    H_5(F) = \mp_1 \Bigl(\pm_2 \norm{p} + \sqrt{z_2}\Bigr) \sqrt{z_1} z_2, \\
    H_6(F) = \norm{p} \sqrt{z_1} \sqrt{z_2} \Bigl(\pm_1 \norm{p} \pm_2 \sqrt{z_1} \pm_3 \sqrt{z_{2}}\Bigr).
\end{gather}
The conditions for a minimum are
\begin{gather}
    (-1)^2 H_5(F) > 0, \qquad
    (-1)^2 H_6(F) > 0,
\end{gather}
where $2$ is the number of constraints.  The conditions for a maximum are
\begin{equation}
    (-1)^3 H_5(F) > 0, \qquad
    (-1)^4 H_6(F) > 0.
\end{equation}

Let us consider the conditions for the minimum:
\begin{gather}
    \mp_1 (\pm_2 \norm{p} + \sqrt{z_2}) > 0, \\
    \pm_1 \norm{p} \pm_2 \sqrt{z_1} \pm_3 \sqrt{z_2} > 0.
\end{gather}
There are four cases
\begin{enumerate}
    \item ${+_1} {+_2} {+_3}$:  in this case the minimum conditions become $\norm{p} + \sqrt{z_2} < 0$ and $\norm{p} + \sqrt{z_1} + \sqrt{z_2} > 0$.  The first condition never holds since $\norm{p} \geq 0$ and $\sqrt{z_2} \geq 0$.
    \item ${+_1} {-_2} {-_3}$: in this case we have $\norm{p} > \sqrt{z_2}$ and $\norm{p} > \sqrt{z_1} + \sqrt{z_2}$.
    \item ${-_1} {+_2} {-_3}$: in this case we have $\norm{p} + \sqrt{z_2} > 0$ and $\sqrt{z_1} > \norm{p} + \sqrt{z_2}$.
    \item ${-_1} {-_2} {+_3}$: in this case we have $\sqrt{z_2} > \norm{p}$ and $\sqrt{z_2} > \norm{p} + \sqrt{z_1}$.
\end{enumerate}

Let us now assume that $\lambda_1 = 0$ or $\lambda_2 = 0$.  Since $\lambda_1 q_1 = \lambda_2 q_2$, then if $\lambda_1 = 0$, either $\lambda_2 = 0$ or $q_2 = 0$.  We will not study the region $q_2 = 0$ anymore since in this case it does not yield a contribution to the integral.  This conclusion should be re-evaluated when studying the case $m_2 = 0$ or in general when the integrals are divergent.  We conclude that when $\lambda_1 = \lambda_2 = 0$ the two analogs of the Landau loop equations become a single equation $p = q_1 + q_2$.  This is consistent with the constraints $q_1^2 = z_1$ and $q_2^2 = z_2$ if the triangle inequalities are satisfied $z_1 + z_2 > \norm{p}$, $\norm{p} + z_1 > z_2$ and $\norm{p} + z_2 > z_1$.

In conclusion, we have
\begin{equation}
    (q_3^2)_{\text{min}} =
    \begin{cases}
        (\norm{p} - \sqrt{z_1} - \sqrt{z_2})^2, &\qquad \norm{p} > \sqrt{z_1} + \sqrt{z_2}, \\
        (\norm{p} - \sqrt{z_1} + \sqrt{z_2})^2, &\qquad \sqrt{z_1} - \sqrt{z_2} > \norm{p}, \\
        (\norm{p} + \sqrt{z_1} - \sqrt{z_2})^2, &\qquad \sqrt{z_2} - \sqrt{z_1} > \norm{p}, \\
        0, &\qquad \text{otherwise}.
    \end{cases}
\end{equation}

For the maximum, we have the following four cases:
\begin{enumerate}
    \item ${+_1} {+_2} {+_3}$: $\norm{p} + \sqrt{z_2} > 0$ and $\norm{p} + \sqrt{z_1} + \sqrt{z_2} > 0$.  These conditions are always satisfied.
    \item ${+_1} {-_2} {-_3}$: $\norm{p} < \sqrt{z_2}$ and $\norm{p} > \sqrt{z_1} + \sqrt{z_2}$.  These conditions are incompatible.
    \item ${-_1} {+_2} {-_3}$: $\norm{p} + \sqrt{z_2} < 0$ and $\sqrt{z_1} > \norm{p} + \sqrt{z_2}$.  These conditions are incompatible.
    \item ${-_1} {-_2} {+_3}$: $\sqrt{z_2} < \norm{p}$ and $\sqrt{z_2} > \norm{p} + \sqrt{z_1}$.  These conditions are incompatible.
\end{enumerate}
It follows that
\begin{equation}
    (q_3^2)_{\text{max}} = (\norm{p} + \sqrt{z_1} + \sqrt{z_2})^2.
\end{equation}
This is actually pretty obvious geometrically.

The innermost integral contains a one-form
\begin{equation}
    \frac{d^2 q_1 d^2 q_2}{d q_1^2 \wedge d q_2^2 \wedge d q_3^2}.
\end{equation}
We can compute this ratio as follows.  We have $q_1 + q_2 + q_3 = p$ and we take $p$ to be constant (or $d p = 0$).  Then we have
\begin{multline}
    \omega = \frac{d^2 q_1 d^2 q_2}{(2 q_1 \cdot d q_1) \wedge (2 q_2 \cdot d q_2) \wedge (2 q_3 \cdot d q_3)} = \\
    \frac{d^2 q_1 d^2 q_2}{-8 \Bigl((q_1 \cdot d q_1) \wedge (q_2 \cdot d q_2) \wedge (q_3 \cdot d q_2) +
    (q_1 \cdot d q_1) \wedge (q_2 \cdot d q_2) \wedge (q_3 \cdot d q_2)\Bigr)} = \\
    \frac{v \cdot d q_1}{8 \epsilon(q_1, q_3) \epsilon(v, q_1)}
\end{multline}
where we have used $d q_3 = -d q_1 - d q_2$.  Then, taking $v = p$ and using
\begin{multline}
    \epsilon(p, q_1) \epsilon(q_2, q_3) =
    \det \begin{pmatrix}
        p \cdot q_2 & p \cdot q_3 \\
        q_1 \cdot q_2 & q_1 \cdot q_3
    \end{pmatrix} = \\
    \det \begin{pmatrix}
        \frac{1}{2} (q_1^2 + q_3^2 - x^2 - y^2) & \frac{1}{2} (y^2 - p^2 - q_3^2) \\
        \frac{1}{2} (y^2 - q_1^2 - q_2^2) & \frac{1}{2} (p^2 + q_2^2 - x^2 - y^2)
    \end{pmatrix},
\end{multline}
where $x^2 = (p - q_1)^2 = (q_2 + q_3)^2$ and $y^2 = (p - q_3)^2 = (q_1 + q_2)^2$.

The variables $x^2$ and $y^2$ are the lengths squared of the diagonals of the quadrilateral whose sides are the vectors $q_1$, $q_2$, $q_3$ and $-p$.

Since the quadrilateral is in a plane, we have that the volume of the simplex it generates vanishes.  In other words, the following Gram determinant should vanish (see also ref.~\cite[lemma~4.1]{MR4092613})
\begin{equation}
    G(p, x, q_3) = \det \begin{pmatrix}
        p^2 & p \cdot x & p \cdot q_3 \\
        p \cdot x & x^2 & x \cdot q_3 \\
        p \cdot q_3 & x \cdot q_3 & q_3^2
    \end{pmatrix} =
    P(u, v),
\end{equation}
where $u = x^2$, $v = y^2$,
\begin{equation}
    P(u, v) = u^2 v + u v^2 + 2 d_{1 1} u v + d_{1 0} u + d_{0 1} v + d_{0 0},
\end{equation}
with
\begin{gather}
    d_{1 1} = -\frac{1}{2} (p^2 + q_1^2 + q_2^2 + q_3^2), \\
    d_{1 0} = (p^2 - q_3^2) (q_1^2 - q_2^2), \\
    d_{0 1} = (p^2 - q_1^2) (q_3^2 - q_2^2), \\
    d_{0 0} = (p^2 - q_1^2 + q_2^2 - q_3^2) (p^2 q_2^2 - q_1^2 q_3^2).
\end{gather}
The polynomial $P$ can be made homogeneous of degree three so it describes an elliptic curve embedded in $\mathbf{P}^2$ with homogeneous coordinates $(u : v : w)$.  However, an alternative compactification, described in sec.~\ref{sec:configurations_quadrilaterals} is more natural.

Using these results we have that
\begin{equation}
    \omega = \frac{d u}{4 \partial_v P} = -\frac{d v}{4 \partial_u P}.
\end{equation}
This one-form can be obtained from the two-form $\frac{u d v d w - v d u d w + w d u d v}{P}$ by taking a residue at $P = 0$.

We can eliminate the variable $v$ by solving the quadratic equation $P = 0$ in $v$ to obtain
\begin{equation}
    v_\pm = \frac{-(u^2 + 2 d_{1 1} u + d_{0 1}) \pm \sqrt{\Delta}}{2 u}
\end{equation}
Using this in $\partial_v P$ we find
\begin{equation}
    \partial_v P = \pm \sqrt{\Delta},
\end{equation}
where
\begin{equation}
    \Delta =
    \Bigl(u - (\norm{p} + \norm{q_1})^2\Bigr)
    \Bigl(u - (\norm{p} - \norm{q_1})^2\Bigr)
    \Bigl(u - (\norm{q_2} + \norm{q_3})^2\Bigr)
    \Bigl(u - (\norm{q_2} - \norm{q_3})^2\Bigr).
\end{equation}

Taking into account the triangle inequality, the domain for $u = x^2$ is as follows.  We have $(\norm{p} - \norm{q_1})^2 \leq x^2$ and $x^2 \leq (\norm{p} + \norm{q_1})^2$.  Similarly, we have $(\norm{q_2} - \norm{q_3})^2 \leq x$ and $x \leq (\norm{q_2} + \norm{q_3})^2$.  Therefore, we have
\begin{equation}
    \max\Bigl((\norm{p} - \norm{q_1})^2, (\norm{q_2} - \norm{q_3})^2\Bigr) \leq u \leq
    \min\Bigl((\norm{p} + \norm{q_1})^2, (\norm{q_2} + \norm{q_3})^2\Bigr).
\end{equation}
In other words, if we sort the roots of $\Delta$, the integral in $u$ is over the middle two roots.  In this interval $\Delta \geq 0$.

Similarly, we have
\begin{equation}
    \partial_u P = \pm \sqrt{\Delta'},
\end{equation}
where
\begin{equation}
    \Delta' = 
    \Bigl(v - (\norm{q_1} + \norm{q_2})^2\Bigr)
    \Bigl(v - (\norm{q_1} - \norm{q_2})^2\Bigr)
    \Bigl(v - (\norm{p} + \norm{q_3})^2\Bigr)
    \Bigl(v - (\norm{p} - \norm{q_3})^2\Bigr).
\end{equation}
As before, taking into account the triangle inequality, the domain for $v = y^2$ is
\begin{equation}
    \max\Bigl((\norm{p} - \norm{q_3})^2, (\norm{q_1} - \norm{q_2})^2\Bigr) \leq u \leq
    \min\Bigl((\norm{p} + \norm{q_3})^2, (\norm{q_1} + \norm{q_2})^2\Bigr).
\end{equation}

We know that
\begin{gather}
    (\norm{p} - \norm{q_3})^2 < (\norm{p} + \norm{q_3})^2, \\
    (\norm{q_1} - \norm{q_2})^2 < (\norm{q_1} + \norm{q_2})^2
\end{gather}
so the possible orderings of these four roots are obtained by shuffling the two sets of ordered roots.  In total there are six possibilities.  The interplay between these conditions and the boundary conditions for integrating over $q_3^2$ yield a large number of regions.

To make progress, we follow a bit of a different route that Cutkosky's.  Instead of integrating over $q_1^2$ and $q_2^2$ first, we integrate over $q_1^2$ and the diagonal $v = y^2$ (see fig.~\ref{fig:quadrilateral_configuration_flipped}).  The triangle inequalities now imply that
\begin{gather}
  q_2^2 \in [(y - \norm{q_1})^2, (y + \norm{q_1})^2], \\
  q_3^2 \in [(y - \norm{p})^2, (y + \norm{p})^2].
\end{gather}
Then the integral becomes
\begin{multline}
  \int_0^\infty \frac {d q_1^2}{q_1^2 + m_1^2}
  \int_0^\infty d v
  \int_{(\sqrt{v} - \norm{q_1})^2}^{(\sqrt{v} + \norm{q_1})^2} \frac {d q_2^2}{q_2^2 + m_2^2} \frac 1 {\sqrt{(v - (\norm{q_1} - \norm{q_2})^2)
      (v - (\norm{q_1} + \norm{q_2})^2)}} \\
  \int_{(\sqrt{v} - \norm{p})^2}^{(\sqrt{v} + \norm{p})^2} \frac {d q_3^2}{q_3^2 + m_3^2} \frac 1 {\sqrt{(v - (\norm{p} - \norm{q_3})^2)
      (v - (\norm{p} + \norm{q_3})^2)}}.
\end{multline}

Next, a short calculation reveals that
\begin{equation}
  (v - (\norm{p} - \norm{q_3})^2)
  (v - (\norm{p} + \norm{q_3})^2) =
  (q_3^2 - (\norm{p} - \sqrt{v})^2)
  (q_3^2 - (\norm{p} + \sqrt{v})^2).
\end{equation}
Then, the integrals over $q_2^2$ and $q_3^2$ can be done as in eq.~\eqref{eq:pole-sqrt-integral} and have the effect of introducing a factor of $\pi$ each and replacing $q_2^2 \to -m_2^2$ and $q_3^2 \to -m_3^2$.

After doing these integrals, we find
\begin{equation}
  \pi^2 \int_0^\infty \frac {d q_1^2}{q_1^2 + m_1^2} \int_0^\infty \frac {d v}{\sqrt{(v - (\norm{p} - i m_3)^2) (v - (\norm{p} + i m_3)^2)
    (v - (\norm{q_1} - i m_2)^2) (v - (\norm{q_1} + i m_2)^2)}}.
\end{equation}
Note that the quantity under the square root is always positive in the integration domain.

Note also that when the ``angular'' integral was the innermost integral it was a \emph{complete} elliptic integral, once we pulled it through the $q_2^2$ and $q_3^2$ integrals it became an \emph{incomplete} elliptic integral (see sec.~\ref{sec:carlson_elliptic} for a discussion of such integrals).

This result can be rewritten in several ways, but the number of integrals can only be reduced at the cost of introducing transcendental functions, such as logarithms.  For example, we can also do the integral over $q_1^2$ which will produce a logarithm and will replace $\norm{q_1} \to i m_1$ in the quartic under the square root.  Similar results can be more quickly obtained by using Feynman parametrization.  The quartic in $v$ can be symmetrically reduced as in eq.~\eqref{eq:symmetric_reduction}.

At the Euclidean pseudo-threshold $\norm{p} = i (m_1 + m_2 - m_3)$ the integration can be done explicitly in terms of dilogarithms as shown in ref.~\cite{Bloch:2013tra}.

Hyperelliptic integrals can also occur, see ref.~\cite{Georgoudis:2015hca}.  In that case a similar analysis applies and the angular integrals should yield a distinguished holomorphic form on the hyperelliptic curve along with a distinguished cycle.

The parametrization of the integral in terms of momenta squared may open up new possibilities for regularization.  The ``angular'' integral, being along a compact real cycle and not meeting any singularities does not itself produce divergences, however this inner integral is the only one affected by dimensional regularization (see the discussion in sec.~\ref{sec:dim-reg}).  Divergences arise from integrals along the variables $q_e^2$.  It is therefore more rational to regularize them instead, since they are producing divergences.  One idea that immediately comes to mind are hard cut-offs (IR and/or UV) in $q_e^2$ in Euclidean signature.  This will undoubtedly make the integrals more complicated, but possibly not much more than dimensional regularization.  We should point out that this type of regularization is better than the usual textbook cutoff regularization, which depends on the choice of loop momentum.

However, the mathematical literature has other types of regularizations which have been applied to polylogarithms and multiple zeta values (see ref.~\cite{goncharov2001multiple} for a longer discussion).  Such regularizations, such as tangential basepoint regularization have already been used in refs.~\cite{Brown2009, Panzer:2014caa}.  The regularizations used in the mathematical literature have been designed to preserve various identities satisfied by quantities which did not require regularization.  Similarly, in physics, we want to preserve various properties satisfied by physical quantities, which are often broken by traditional regularization choices.

\section{Configurations of quadrilaterals as elliptic curves}
\label{sec:configurations_quadrilaterals}

It will prove convenient to compactify and complexify the integration domain.  This is actually necessary for applying mathematical theorems such as those of refs.~\cite{BSMF_1959__87__81_0, MR214101, pham, pham2011singularities} and also refs.~\cite{Hwa:102287, Hannesdottir:2022xki} for reviews.  The compactification is essential if we want to study second type or mixed second type singularities.  For the variables $q_e^2$ an obvious choice of compactification is $\mathbf{RP}^1$ with complexification $\mathbf{CP}^1$ (see ref.~\cite{MR4092613} where the same compactification is used).

It may happen that, after complexification, the ``angular'' variables in Cutkosky's terminology (see sec.~\ref{sec:cutkosky}) do not parametrize a compact space.  Sometimes, as in the case of the bubble integral in three dimensions (see sec.~\ref{sec:bubble-3d}), a compactification can be performed at the cost of introducing a pole for the innermost differential form.  We should note that this rather natural (partial) compactification does not seem to have been discussed in the physics literature before.  The type of compactifications that have been considered, see ref.~\cite[p.~107]{Hwa:102287} involve representing the complexified compactified Minkowski space as a quadric in $\mathbf{P}^5$, a compactification familiar to twistor theorists.  Here, instead, we are proposing to use a compactification to a product of $\mathbf{P}^1$, times an ad hoc compactification for the ``angular'' variety.  Curiously, an embedding in a product of $\mathbf{P}^1$ was discussed in a different context in ref.~\cite{Vergu:2020uur} but there the interpretation of the coordinate on $\mathbf{P}^1$ was different from here.

In the compactification the integration path $q_e^2 \in (-\infty, \infty)$ is closed since we have a single point at infinity.  Indeed, on the complex projective line $\mathbf{CP}^1$ or the Riemann sphere we can choose coordinates so that the origin is at the North pole and the infinity is at the South pole.  Then the integration contour $(-\infty, \infty)$ is along a meridian.

\begin{figure}
    \centering
    \includegraphics{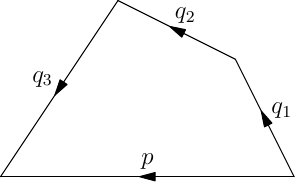}
    \caption{A quadrilateral configuration, which corresponds to a kinematic point in the sunrise integral.}
    \label{fig:quadrilateral_configuration}
\end{figure}

\begin{figure}
    \centering
\includegraphics{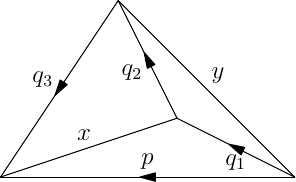}
    \caption{A quadrilateral configuration, together with diagonals, which has the same lengths of the sides, and same length for one of the diagonals as in fig.~\ref{fig:quadrilateral_configuration}, while the other diagonal has a different length.}
  \label{fig:quadrilateral_configuration_flipped}
\end{figure}

The interpretation of quadrilateral configurations as points of elliptic curve was described in ref.~\cite{MR4092613}.  As explained in this reference, there are two moduli spaces of quadrilaterals; oriented and unoriented.  For our purposes the unoriented moduli space is relevant.  The oriented moduli space of quadrilaterals makes sense over the real numbers only.

In order to obtain an elliptic curve, we need to compactify the space, which involves taking the lengths of the sides of the quadrilateral to be valued in $\mathbf{P}^1$.

The dual space parametrization used here goes back the initial studies of Landau singularities in refs.~\cite{Karplus:1958zz, Karplus:1959zz, landau1959, OKUN1960261}.  This representation is really useful since some non-obvious algebraic properties have simple geometric causes.  An equation for the elliptic curve can be obtained by setting to zero the volume of a (degenerate) tetrahedron in fig.~\ref{fig:quadrilateral_configuration_flipped}.  This yields an equation in $u = x^2$ and $v = y^2$ of bi-degree $(2, 2)$ and is, coincidentally, also naturally embedded in $\mathbf{P}^1 \times \mathbf{P}^1$.

\section{A peek at integrals in dimensional regularization}
\label{sec:dim-reg}

In dimensional regularization the Cutkosky representation is usually called Baikov representation (see ref.~\cite{Baikov_1996} for the original paper and ref.~\cite{Zhang:2016kfo} for an introduction).  This has a loop-by-loop version worked out in ref.~\cite{Frellesvig:2021vdl}.

Let us now do a sample computation in dimensional regularization.  We will first attempt a simple integral, the massless bubble in dimension $d = 4 - 2 \epsilon$.  Our computation will not use the Wick rotation, which is less natural for massless particles since it clashes with the on-shell conditions.  The computation is more complicated than the textbook computation using Feynman parameters and Wick rotation.

We have
\begin{equation}
  \int \frac {d^d q_1^2}{q_1^2 q_2^2} =
  \int_{-\infty}^\infty \frac {d q_1^2}{q_1^2} \Bigl(\int_{-\infty}^{a_2} + \int_{b_2}^\infty\Bigr) \frac {d q_2^2}{q_2^2} \int \frac {d^d q_1}{d q_1^2 \wedge d q_2^2}.
\end{equation}

As before, we have $a_2 = (\norm{p} - \norm{q_1})^2$ and $b_2 = (\norm{p} + \norm{q_1})^2$.  Recall that $\norm{p} = \sqrt{p^2}$ and if $p^2 < 0$, then $\norm{p} \in i \mathbb{R}$.  In this case, when the boundaries of integration are not real, the inner integral is over the full real line so we replace
\begin{equation}
  \Bigl(\int_{-\infty}^{a_2} + \int_{b_2}^\infty\Bigr) \frac {d q_2^2}{q_2^2} \to
  \int_{-\infty}^\infty \frac {d q_2^2}{q_2^2}
\end{equation}
above.  In the following we will do the computation in the region $p^2 < 0$, sometimes called Euclidean region.

We have
\begin{equation}
  \frac {d^d q_1}{d q_1^2 \wedge d q_2^2} =
  -\frac {d^d q_1}{4 (q_1 \cdot d q_1) \wedge (q_2 \cdot d q_1)} =
  -\frac 1 4 \frac {d q_1^2 \wedge \cdots \wedge d q_1^{d - 1}}{p^0 q_1^1 - p^1 q_1^0}.
\end{equation}

Next, we use the fact that
\begin{equation}
  (p^0 q_1^1 - p^1 q_1^0)^2 =
  (p^0 q_1^0 - p^1 q_1^1) - ((p^0)^2 - (p^1)^2) ((q_1^0)^2 - (q_1^1)^2).
\end{equation}
We pick $p$ to have components only along directions $0$ and $1$ so $p^2 = (p^0)^2 - (p^1)^2$.  Then, denoting $q_1^2 = z_1$ and $q_2^2 = z_2$, we have
\begin{gather}
  (q_1^0)^2 - (q_1^1)^2 = z_1 + (q_1^2)^2 + \cdots (q_1^{d-1})^2, \\
  (p^0 - q_1^0)^2 - (p^1 - q_1^1)^2 = z_2 + (q_1^2)^2 + \cdots (q_1^{d-1})^2.
\end{gather}
Subtracting these equalities we find $p^2 - 2 (p^0 q_1^0 - p^1 q_1^1) = z_2 - z_1$.  Using this we find
\begin{equation}
  \label{eq:dim-reg-jacobian}
  (p^0 q_1^1 - p^1 q_1^0)^2 =
  \frac 1 4 (z_2 - z_1 - p^2)^2 - p^2 (z_1 + \rho^2),
\end{equation}
where we used $\rho^2 = (q_1^2)^2 + \cdots (q_1^{d-1})^2$.

In conclusion, we have
\begin{equation}
  \label{eq:inner-measure}
  \frac {d^d q_1}{d q_1^2 \wedge d q_2^2} =
  \pm \frac {\Omega_{d - 3} \rho^{d - 3} d \rho}{\sqrt{\frac 1 4 (z_1^2 + z_2^2 + (p^2)^2 - 2 z_1 p^2 - 2 z_2 p^2 - 2 z_1 z_2) - p^2 \rho^2}},
\end{equation}
where $\Omega_{d - 3}$ is the measure on the unit $d-3$-dimensional sphere.  In eq.~\eqref{eq:dim-reg-jacobian} the right-hand side is positive since the left-hand side is a square of a real quantity.  Hence, the quantity under square root in eq.~\eqref{eq:inner-measure} is also positive and taking the square root poses no problem.

In general we have $\Omega_d = \frac {2 \pi^{\frac {d + 1} 2}}{\Gamma(\frac {d + 1} 2)}$ so the angular integral is easy while the radial integral can be done in terms of an Euler beta function.  After integration (for $\rho \in [0, \infty)$) we find
\begin{equation}
  \pm \frac {2 \pi^{\frac {d - 2} 2}}{\Gamma\bigl(\frac {d - 2} 2\bigr)} \frac {\Gamma\bigl(\frac {d - 2} 2\bigr) \Gamma\bigl(\frac {3 - d} 2\bigr)}{2 \sqrt{\pi}} (-p^2)^{-\frac {d - 2} 2} \Bigl(\frac 1 4 (z_1^2 + z_2^2 + (p^2)^2 - 2 z_1 p^2 - 2 z_2 p^2 - 2 z_1 z_2)\Bigr)^{\frac {d - 3} 2},
\end{equation}
for $\Re d \in (2, 3)$.  This form of the integral can be analytically continued in $d$.

The kinematic dependence of this problem is so simple that now we can do a change of variable $z_1 \to -p^2 z_1$ and $z_2 \to -p^2 z_2$ so that the dependence on $p^2$ can be extracted out of the integral and what remains to compute is a $d$-dependent prefactor.  Upon performing this change of variable we can pull out a factor $(-p^2)^{\frac d 2 - 2}$, which is in fact dictated by dimensional analysis.

If instead we take $p^2 > 0$, then there is a finite upper bound on $\rho$.  Doing the $\rho$ integral we obtain
\begin{equation}
  \pm \frac {2 \pi^{\frac {d - 2} 2}}{\Gamma\bigl(\frac {d - 2} 2\bigr)} \frac {\sqrt{\pi} \Gamma\bigl(\frac {d - 2} 2\bigr)}{2 \Gamma\bigl(\frac {d - 1} 2\bigr)} (p^2)^{-\frac {d - 2} 2} \Bigl(\frac 1 4 (z_1^2 + z_2^2 + (p^2)^2 - 2 z_1 p^2 - 2 z_2 p^2 - 2 z_1 z_2)\Bigr)^{\frac {d - 3} 2},
\end{equation}
for $\Re d \in (2, \infty)$.

Keeping only the $z_1$ and $z_2$ dependent part, we have
\begin{multline}
  \Bigl(\int_{-\infty}^0 \frac {d z_1}{z_1} \bigl(\int_{-\infty}^{a_2(z_1)} + \int_{b_2(z_1)}^\infty\bigr) \frac {d z_2}{z_2} +
  \int_0^\infty \frac {d z_1}{z_1} \int_{-\infty}^\infty \frac {d z_2}{z_2}\Bigr) \\
  \bigl(z_1^2 + z_2^2 + 1 + 2 z_1 + 2 z_2 - 2 z_1 z_2\bigr)^{\frac {d - 3} 2}.
\end{multline}
The integrals over $z_1$ and $z_2$ should be understood in the sense of the Feynman $i \varepsilon$ prescription.  Rotating the $z_1$ contour in the second integral above (as we have done in sec.~\ref{sec:bubble-lorentzian}) we obtain
\begin{equation}
  \label{eq:z-integrations}
  -\int_{-\infty}^0 \frac {d z_1}{z_1} \int_{a_2(z_1)}^{b_2(z_1)} \frac {d z_2}{z_2} \Bigl((z_2 - a_2(z_1)) (z_2 - b_2(z_1))\Bigr)^{\frac {d - 3} 2},
\end{equation}
where $a_2(z_1) = -(1 + \sqrt{-z_1})^2$ and $b_2(z_1) = -(1 - \sqrt{-z_1})^2$, so $a_2(z_1) < b_2(z_1) < 0$.  Hence, the integration region does not contain the pole at $z_2 = 0$ and Feynman $i \varepsilon$ are not required to make sense of the integral.

Consider the following integral
\begin{equation}
  \int_a^b \frac {d z} z (z - a)^{\frac 1 2} (z - b)^{\frac 1 2},
\end{equation}
for $a < b < 0$, which arises from the integral above upon setting $d = 4$.  We can write this integral as half the integral around the cut between $a$ and $b$.  This integral is $2 \pi i$ times the residues at $z = 0$ and $z = \infty$.  We have $\operatorname{res}_0 = \sqrt{-a} \sqrt{-b}$.  At infinity we make a change of coordinates $z = w^{-1}$ so we get
\begin{equation}
  \frac {d z} z (z - a)^{\frac 1 2} (z - b)^{\frac 1 2} =
  -\frac {d w}{w^2} (1 - w a)^{\frac 1 2} (1 - w b)^{\frac 1 2} =
  -\frac {d w}{w^2} + \frac {a d w}{2 w} + \frac {b d w}{2 w} + \cdots,
\end{equation}
so $\operatorname{res}_\infty = \frac 1 2 (a + b)$.  Hence, the integral is
\begin{equation}
  \int_a^b \frac {d z} z (z - a)^{\frac 1 2} (z - b)^{\frac 1 2} =
  \frac {\pi i} 2 (a + b + \sqrt{a b}).
\end{equation}
Using this expression for $a = a_2(z_1)$ and $b = b_2(z_1)$ we find
\begin{equation}
  \int_{a_2(z_1)}^{b_2(z_1)} \frac {d z_2}{z_2} (z_2 - a_2(z_1))^{\frac 1 2} (b_2(z_1) - z_2)^{\frac 1 2} = \frac \pi 2 \begin{cases}
    4 z_1, &\qquad \text{if $-1 < z_1 < 0$}, \\
    -4, &\qquad \text{if $z_1 < -1$}.
    \end{cases}
\end{equation}

Plugging this into the $z_1$ integral, we see that it is convergent at $z_1 = 0$ but it diverges logarithmically at $z_1 \to \infty$.  The first region can be thought as an IR region while the second as an UV region so our integral is IR-finite but UV-divergent.  Interestingly, the value $z_1 = -1$ provides a natural boundary between these two regions so one can canonically separate the integral into an IR and a UV region, to be studied separately.

If we do not set $d = 4$, the integral over $z_2$ in eq.~\eqref{eq:z-integrations} is a hypergeometric integral so it can be computed explicitly.
\begin{multline}
  \int_a^b \frac {d x} x (x - a)^e (b - x)^e =
  \frac {(b - a)^{2 e + 1}} a \int_0^1 d y y^e (1 - y)^e (1 - \zeta y)^{-1} = \\
  \frac {(b - a)^{2 e + 1}} a \frac {\Gamma(e + 1)^2}{\Gamma(2 e + 2)} \; {}_2F_{1}(1, e + 1; 2 e + 2; \zeta),
\end{multline}
where $\zeta = 1 - \frac b a$.  We want to apply this for $e = \frac {d - 3} 2$.  However, we are only interested in computing the expansion around $\epsilon \to 0$.

Next, we will seek to establish that this hypergeometric integral has the following behavior
\begin{multline}
  \int_{a_2(z_1)}^{b_2(z_1)} \frac {d z_2}{z_2} \Bigl((z_2 - a_2(z_1)) (z_2 - b_2(z_1))\Bigr)^{\frac {d - 3} 2} \\ \sim
  -2 \pi \begin{cases}
    (-z_1)^{c_0 (d - 4) + 1} + \mathcal{O}(d - 4), &\qquad \text{when $z_1 \to 0$}, \\
    (-z_1)^{c_\infty (d - 4)} + \mathcal{O}(d - 4), &\qquad \text{when $z_1 \to \infty$},
  \end{cases}
\end{multline}
where $c_0$ and $c_\infty$ are constants to be determined.

Then we can obtain the leading behavior in $\frac 1 {d - 4}$ by using
\begin{equation}
  \int_{-1}^0 \frac {d z_1}{z_1} (-z_1)^{c_0 (d - 4) + 1} =
  \left.\frac {(-z_1)^{c_0 (d - 4) + 1}}{c_0 (d - 4) + 1}\right\rvert_{z_1 = -1}^{z_1 = 0} =
  -\frac 1 {c_0 (d - 4) + 1},
\end{equation}
if $\Re (c_0 (d - 4)) > -1$.

Similarly, we have
\begin{equation}
  \int_{-\infty}^{-1} \frac {d z_1}{z_1} (-z_1)^{c_\infty (d - 4)} =
  \left.\frac {(-z_1)^{c_\infty (d - 4)}}{c_\infty (d - 4)}\right\rvert_{z_1 = -\infty}^{z_1 = -1} =
  \frac 1 {c_\infty (d - 4)},
\end{equation}
if $\Re (c_\infty (d - 4)) < 0$.

We have
\begin{equation}
  \zeta = 1 - \frac {b_2(z_1)}{a_2(z_1)} =
  \frac {4 \sqrt{-z_1}}{(1 + \sqrt{-z_1})^2},
\end{equation}
and
\begin{equation}
  b - a = -(1 - \sqrt{-z_1})^2 + (1 + \sqrt{-z_1})^2 = 4 \sqrt{-z_1}.
\end{equation}
Plugging back into the expression terms of hypergeometric functions we have
\begin{equation}
  \frac {(b - a)^{2 e + 1}}{a} =
  \frac {(4 \sqrt{-z_1})^{2 e + 1}}{-(1 + \sqrt{-z_1})^2} =
  -4^{2 e + 1} \times
  \begin{cases}
    (-z_1)^{e + \frac 1 2}, &\qquad z_1 \nearrow 0, \\
    (-z_1)^{e - \frac 1 2}, &\qquad z_1 \searrow -\infty.
  \end{cases}
\end{equation}
Recall that $e = \frac {d - 3} 2 = \frac 1 2 - \epsilon$.  Hence, $c_0 = c_\infty = \frac 1 2$.

This was an woeful derivation, which we hope to improve later.  What is needed is a way to systematically expand around $z_1 = 0$ in the first region and $z_1 \to -\infty$ in the second region.

Finally let us briefly discuss the massless box integral.  We have
\begin{equation}
  \int \frac {d^D k}{k^2 (k + p_2)^2 (k + p_{23})^2 (k - p_1)^2}.
\end{equation}
Let us put this integral in Cutkosky form.  The range of $k^2$ is $\mathbb{R}$.  The range of $(k + p_2)^2$ can be determined by extremizing $(k + p_2)^2$ at fixed $p_2$ and subject to the constraint that $k^2$ is fixed at $k^2 = z_1$.  Using Lagrange multipliers we find
\[
  \frac \partial {\partial k} \bigl((p + k_2)^2 + \alpha_1 k^2\bigr) = 2 (k + p_2 + \alpha_1 k) = 0.
\]
Since $p_2^2 = 0$ and $z_1 = k^2 \neq 0$ we have that an extremum is never realized.  Hence, the range of $(k + p_2)^2$ is also $\mathbb{R}$.  Note that this is not what happens in Cutkosky's approach of proving his theorem.  It remains to be seen if and how his proof would have to be modified to cover this case.

The range of $(k + p_{23})^2$ is determined in a similar way.  Using the same idea of Lagrange multipliers we find the equation
\[
  (1 + \alpha_1 + \alpha_2) k = -p_{23} - \alpha_2 p_3.
\]
The compatibility condition is the existence of a tetrahedron with sides $k$, $k + p_2$, $k + p_{23}$, $p_3$, $p_{14}$ and $p_2$ and the extrema arise when the tetrahedron becomes degenerate which is when its volume vanishes.  Using the Cayley-Manger formula for the volume of the tetrahedron one can solve for the stationary point of $z_3 = (k + p_{23})^2$ and we find a unique solution $z_3 = -\frac {(p_{23}^2 - z_1 + z_2) z_2}{z_1 - z_2}$.  It is remarkable that there is a unique solution which is another departure from the case analyzed by Cutkosky.  We defer a more detailed study of the nature of this stationary point.

The extrema of $(k - p_1)^2$ can be analyzed similarly and in that case one finds two solutions, as usual.

The remaining integrals can be written as
\begin{multline}
  \frac {d^D k}{d k^2 \wedge d (k + p_2)^2 \wedge d (k + p_{23})^2 \wedge d (k - p_1)^2} = \\
  \frac {d^D k}{16 (k \cdot d k) \wedge ((k + p_2) \cdot d k) \wedge ((k + p_{23}) \cdot d k) \wedge ((k - p_1) \cdot d k)} = \\
  \frac {d^{D - 4} k_\perp}{16 \epsilon(k_\parallel, (k + p_2)_\parallel, (k + p_{23})_\parallel, (k - p_1)_\parallel)}.
\end{multline}

Next, we will use the following identity
\begin{multline}
  -\epsilon(k, k + p_2, k + p_{23}, k - p_1)^2 = \\
  \det \begin{pmatrix}
    k^2 & k \cdot (k + p_2) & k \cdot (k + p_{23}) & k \cdot (k - p_1) \\
    k \cdot (k + p_2) & (k + p_2)^2 & (k + p_2) \cdot (k + p_{23}) & (k + p_2) \cdot (k - p_1) \\
    k \cdot (k + p_{23}) & (k + p_2) \cdot (k + p_{23}) & (k + p_{23})^2 & (k + p_{23}) \cdot (k - p_1) \\
    k \cdot (k - p_1) & (k + p_2) \cdot (k - p_1) & (k + p_{23}) \cdot (k - p_1) & (k - p_1)^2
  \end{pmatrix} = \\
  -\epsilon(k_\parallel, (k + p_2)_\parallel, (k + p_{23})_\parallel, (k - p_1)_\parallel)^2 -
  \frac 1 4 (p_{12}^2 + p_{23}^2) p_{12}^2 p_{23}^2 \rho^2,
\end{multline}
where we have used the fact that every element in the Gram matrix above can be written by replacing $k \to k_\parallel$ and adding $\rho^2 = k_\perp^2$.  This allows us to compute the Jacobian as
\begin{multline}
  \epsilon(k_\parallel, (k + p_2)_\parallel, (k + p_{23})_\parallel, (k - p_1)_\parallel) = \\
  \sqrt{\epsilon(k, k + p_2, k + p_{23}, k - p_1)^2 - \frac 1 4 (p_{12}^2 + p_{23}^2) p_{12}^2 p_{23}^2 \rho^2},
\end{multline}
where $\epsilon(k, k + p_2, k + p_{23}, k - p_1)^2$ is a relatively complicated polynomial in $z_1, \dotsc, z_4$ and $p_{12}^2$ and $p_{23}^2$, which will not write explicitly, but which can be computed straightforwardly by expanding the Gram determinant above.  Somewhat unexpectedly the dependence on $z$ cancels from the coefficient of $\rho^2$.

Next, the strategy is as in the case of the bubble integral.  We write the integral over $k_\perp$ as $d^{D - 4} k_\perp = \rho^{D - 5} d \rho d \Omega_{d - 5}$ and perform both the angular and the radial integrals.  The integral over $z_4$ is again a hypergeometric integral, so if one wants to do the integrals to all orders in $\epsilon$ one would need to use identities for integrals with hypergeometric function integrals.  Interestingly, the full answer for the box can be computed exactly in terms of hypergeometric functions, see ref.~\cite{Brandhuber_2008}.

\appendix

\section{Useful integrals}
\label{sec:useful_integrals}

Consider the integral
\begin{equation}
  \int_a^b \frac {d x}{(x + c) \sqrt{(x - \alpha) (\beta - x)}},
\end{equation}
where $a < b$, $\alpha < \beta$, the interval $(a, b)$ is included in the interval $(\alpha, \beta)$ so the square root is always positive in the integration region and $-c \not\in (a, b)$.

To compute this integral we define a curve $y^2 = (x - \alpha) (\beta - x)$ which can also be written
\begin{equation}
  (x - \frac {\alpha + \beta} 2)^2 + y^2 = (\frac {\alpha - \beta} 2)^2.
\end{equation}
This curve can be parametrized rationally by
\begin{gather}
  x = \frac {\alpha + \beta} 2 + \frac {\alpha - \beta}{4 i} (t - t^{-1}), \\
  y = \frac {\alpha - \beta} 4 (t + t^{-1}).
\end{gather}

In this new coordinate we have
\begin{equation}
  \omega = \frac {d x}{(x + c) \sqrt{(x - \alpha) (\beta - x)}} =
  \frac 1 {2 \sqrt{-(\alpha + c)(c + \beta)}} d \log \frac {t - t^+}{t - t^-},
\end{equation}
where
\begin{equation}
  t^\pm = \frac {-i (\alpha + \beta + 2 c) \pm 2 \sqrt{-(\alpha + c)(c + \beta)}}{\alpha - \beta}.
\end{equation}

Then we solve for $t_i$ corresponding to $x = a$ and for $t_f$ corresponding to $x = b$.  In both cases there are two solutions and we pick the solution for which $y > 0$.  We therefore have
\begin{gather}
  t_i = \frac {i (2 a - \alpha - \beta) + 2 \sqrt{(a - \alpha) (\beta - a)}}{\alpha - \beta}, \\
  t_f = \frac {i (2 b - \alpha - \beta) + 2 \sqrt{(b - \alpha) (\beta - b)}}{\alpha - \beta}.
\end{gather}
Using these equations we find
\begin{equation}
  \int_a^b \frac {d x}{(x + c) \sqrt{(x - \alpha) (\beta - x)}} =
  \frac 1 {2 \sqrt{-(\alpha + c)(c + \beta)}} \log \frac {(t_f - t^+) (t_i - t^-)}{(t_f - t^-) (t_i - t^+)},
\end{equation}
where we have taken $\alpha < a < b < \beta$ and $-c \not\in (\alpha, \beta)$.

An alternative form for the answer is
\begin{multline}
  \int_a^b \frac {d x}{(x + c) \sqrt{(x - \alpha) (\beta - x)}} =
  \frac 2 {\sqrt{(\alpha + c)(\beta + c)}} \times \\\Biggl(
  \arctan \sqrt{\frac {(a - \beta) (\alpha + c)}{(\alpha - a) (\beta + c)}} -
  \arctan \sqrt{\frac {(b - \beta) (\alpha + c)}{(\alpha - b) (\beta + c)}}
  \Biggr).
\end{multline}
Notice the appearance of square roots of cross-ratios of points $a, \beta, \alpha, -c$ and $b, \beta, \alpha, -c$.  Interestingly, square roots of cross-ratios have appeared in the recent work of Rudenko (see ref.~\cite{rudenko2022goncharov}).

We have
\begin{equation}
  \arctan x = \frac 1 {2 i} \log \frac {1 + i x}{1 - i x}.
\end{equation}
Then, $\arctan \sqrt{x}$ has a square root branch cut at $x = 0$ and a logarithmic branch cut at $x = -1$.  The logarithmic branch points are obtained by solving
\begin{equation}
  \frac {(a - \beta)(\alpha + c)}{(\alpha - a)(\beta + c)} = -1
\end{equation}
we have $\alpha = \beta$ or $a = -c$.  Similarly, from the second arctangent function, we have logarithmic branch points when
\begin{equation}
  \frac {(b - \beta)(\alpha + c)}{(\alpha - b)(\beta + c)} = -1,
\end{equation}
which implies $\alpha = \beta$ or $b = -c$.

If the roots of the quadratic polynomial under the square root are complex, then we have the integral
\begin{equation}
  \int_a^b \frac {d x}{(x + c) \sqrt{(x - z)(x - \bar{z})}},
\end{equation}
where $-c < a < b$ and $z = u + i v$, $\bar{z} = u - i v$ for $u, v \in \mathbb{R}$ and $v \geq 0$.  The calculation of this integral is classical.  We first define a curve $y^2 = (x - u)^2 + v^2$, which can be written as $(y - x + u)(y + x - u) = v^2$.  This curve can be parametrized by
\begin{equation}
  x = u + \frac 1 2 (\frac {v^2} t - t), \qquad
  y = \frac 1 2 (t + \frac {v^2} t).
\end{equation}
We have
\begin{equation}
  \omega = \frac {d x}{(x + c) \sqrt{(x - z)(x - \bar{z})}} =
  \frac {2 d t}{t^2 - 2 (u + c) t - v^2}.
\end{equation}
After partial fractioning this becomes
\begin{equation}
  \omega = \frac 1 {\sqrt{(u + c)^2 + v^2}} d \log \frac {t - t_+}{t - t_-},
\end{equation}
where
\begin{equation}
  t_\pm = u + c \pm \sqrt{(u + c)^2 + v^2}.
\end{equation}

The bounds of integration can be determined using $x(t_i) = a$ and $x(t_f) = b$.  There are two solutions in each case but we keep the one for which $y > 0$.  In the end we find
\begin{gather}
  t_i = u - a + \sqrt{(u - a)^2 + v^2}, \qquad
  t_f = u - b + \sqrt{(u - b)^2 + v^2}.
\end{gather}

Finally the integral is
\begin{multline}
  \label{eq:known-integral}
  \int_a^b \omega =
  \frac 1 {\sqrt{(u + c)^2 + v^2}} \Bigl[
  \log \bigl(\frac {-b - c + \sqrt{(b - u)^2 + v^2} - \sqrt{(u + c)^2 + v^2}}{-b - c + \sqrt{(b - u)^2 + v^2} + \sqrt{(u + c)^2 + v^2}}\bigr) \\-
  \log \bigl(\frac {-a - c + \sqrt{(a - u)^2 + v^2} - \sqrt{(u + c)^2 + v^2}}{-a - c + \sqrt{(a - u)^2 + v^2} + \sqrt{(u + c)^2 + v^2}}\bigr)
  \Bigr].
\end{multline}

Let us now study the singularities of this function.  We have logarithmic singularities when the arguments of the logarithms become zero or infinity.  For the first logarithm this happens when
\begin{equation}
  \bigl(-b - c + \sqrt{(b - u)^2 + v^2} - \sqrt{(u + c)^2 + v^2}\bigr) \bigl(-b - c + \sqrt{(b - u)^2 + v^2} + \sqrt{(u + c)^2 + v^2}\bigr) = 0.
\end{equation}
This condition is satisfied if $b + c = 0$ or $v = 0$ while $u \in [a, b]$.  In the second case the contour $[a, b]$ is pinched.  A similar conclusion holds for the second logarithm with $a$ substituted for $b$.

\begin{figure}
    \centering
    \includegraphics{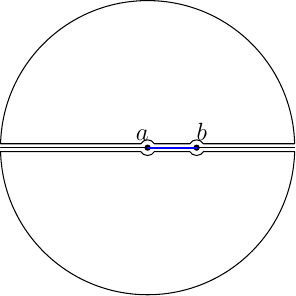}
    \caption{Two contours along the real axis with small positive and negative imaginary parts.  The contours avoid the singularities at $a$ and $b$ along the real axis and they close by large semicircles in the upper and lower half-planes.}
    \label{fig:sqrt-cut3}
\end{figure}

The usual branch cut prescription for $\sqrt{(x - a)(x - b)}$ is that if the quantity under square root has a negative real part and a small positive imaginary part then the argument of the square root is $\frac{i \pi}{2}$.  If instead it has a small negative imaginary part then the argument of the square root is $-\frac{i \pi}{2}$.  From this point of view $\sqrt{(x - a)(x - b)}$ is badly behaved since we have $(x + i \epsilon - a)(x + i \epsilon - b) = (x - a)(x - b) - i \epsilon (a + b - 2 x) + \mathcal{O}(\epsilon^2)$.  Here we assume $a, b, x, \epsilon \in \mathbb{R}$ and for definiteness $a < b$.  Then $\Re \bigl((x + i \epsilon - a)(x + i \epsilon - b)\bigr) < 0$ if $x \in (a, b)$.  We also have $\Im \bigl((x + i \epsilon - a)(x + i \epsilon - b)\bigr) = - \epsilon (a + b - 2 x)$.  This changes sign when $x = \frac{a + b}{2}$.

It is more convenient to instead use $\sqrt{x - a} \sqrt{x - b}$.  For $x > b$ there are no branch cuts.  For $x \in (a, b)$ there is a square root branch cut arising from $\sqrt{x - b}$.  Finally, for $x < a$ there is no branch cut as can be shown by a short calculation.  One can also choose the complementary cut, by considering the function $\sqrt{x - a} \sqrt{b - x}$ instead.  Indeed, as long as $x \in (a, b)$ the function is continuous from above and below in the complex plane while along $x < a$ and $x > b$ we have the usual square root branch cuts of $\sqrt{x - a}$ and $\sqrt{b - x}$, respectively.

The contours in fig.~\ref{fig:sqrt-cut3} are very useful in computing integrals of type
\begin{equation}
    \int_a^b \frac{d x}{f(x) \sqrt{x - a} \sqrt{x - b}},
\end{equation}
where $f$ does not have any singularities along the real axis.  Then, we have
\begin{equation}
    \int_{-\infty + i \delta}^{a - \sqrt{\epsilon^2 - \delta^2} + i \delta} +
    \int_{a - \sqrt{\epsilon^2 - \delta^2} + i \delta}^{a + \sqrt{\epsilon^2 - \delta^2} + i \delta} +
    \int_{a + \sqrt{\epsilon^2 - \delta^2} + i \delta}^{b - \sqrt{\epsilon^2 - \delta^2} + i \delta} +
    \int_{b - \sqrt{\epsilon^2 - \delta^2} + i \delta}^{b + \sqrt{\epsilon^2 - \delta^2} + i \delta} +
    \int_{b + \sqrt{\epsilon^2 - \delta^2} + i \delta}^{\infty + i \epsilon},
\end{equation}
which is a decomposition of the horizontal contour in the upper plane in fig.~\ref{fig:sqrt-cut3}.  This contour has several portions.  The horizontal portions are displaced by $\delta > 0$ in the upper half plane while the arcs of circle have radius $\epsilon > \delta$.  If we subtract from this path the analogous path in the lower half plane, the linear sections going to $\pm \infty$ cancel, while the section along the cut doubles since it receives a contribution from the contour in the upper half plane and a negative contribution (due to the branch cut) along a contour going in the opposite direction.

In practice $f$ is often such that the integrals along the small circles of radius $\epsilon$ vanish in the limit $\epsilon \to 0$ (which also implies $\delta \to 0$).  Then one can compute the integral by an application of Cauchy's theorem, by closing the contours at infinity.  If $f$ has branch cuts, we can arrange so the contours go along the branch cuts without crossing them.  This produces contributions to the integral from Cauchy's theorem.  Finally, we need to take into account the poles, including a potential pole at infinity.

If instead we want to compute an integral of type
\begin{equation}
    \int_{(-\infty, a) \cup (b, \infty)} \frac{d x}{f(x) \sqrt{x - a} \sqrt{x - b}},
\end{equation}
we can follow a similar recipe, except that now we \emph{add} the contributions of the two horizontal contours instead of subtracting them.

\section{Carlson elliptic integrals}
\label{sec:carlson_elliptic}

In ref.~\cite[thm.~2.2]{MR1737495} the following duplication theorem was proved.  If $z_1, z_2, z_3 \in \mathbb{C} \setminus (-\infty, 0)$ and at most one of them is zero, then we have
\begin{equation}
    \int_0^\infty \frac{d t}{\prod_{i = 1}^3 \sqrt{t + z_i}} =
    2 \int_0^\infty \frac{d u}{\prod_{i = 1}^3 \sqrt{u + z_i + \lambda}},
\end{equation}
where
\begin{equation}
    \lambda = \sqrt{z_1} \sqrt{z_2} + \sqrt{z_1} \sqrt{z_3} + \sqrt{z_2} \sqrt{z_3}
\end{equation}
and where the square roots have phases in the right half-plane.  The duplication theorem has to do with an isogeny of the elliptic curve.  See refs.~\cite{Adams:2017ejb, Bogner:2019lfa, Frellesvig:2021vdl} for discussions of isogenies in connection with Feynman integrals.

This can be proved by making a change of variable
\begin{equation}
    u(t) = t + \sqrt{t + z_1} \sqrt{t + z_2} + \sqrt{t + z_1} \sqrt{t + z_3} + \sqrt{t + z_2} \sqrt{t + z_3} - \lambda.
\end{equation}
Then we have
\begin{equation}
    2 \frac{d u}{d t} = 2 + \frac{\sqrt{t + z_2}}{\sqrt{t + z_1}} +
    \frac{\sqrt{t + z_1}}{\sqrt{t + z_2}} +
    \frac{\sqrt{t + z_3}}{\sqrt{t + z_1}} +
    \frac{\sqrt{t + z_1}}{\sqrt{t + z_3}} +
    \frac{\sqrt{t + z_2}}{\sqrt{t + z_3}} +
    \frac{\sqrt{t + z_3}}{\sqrt{t + z_2}}.
\end{equation}
Next, we have
\begin{equation}
    u + z_1 + \lambda = (\sqrt{t + z_1} + \sqrt{t + z_2}) (\sqrt{t + z_1} + \sqrt{t + z_3})
\end{equation}
and permutations.  Hence,
\begin{multline}
    \sqrt{(u + z_1 + \lambda) (u + z_2 + \lambda) (u + z_3 + \lambda)} = \\
    (\sqrt{u + z_1} + \sqrt{u + z_2})
    (\sqrt{u + z_1} + \sqrt{u + z_3})
    (\sqrt{u + z_2} + \sqrt{u + z_3}).
\end{multline}
Finally, simple algebra shows that
\begin{equation}
    \frac{d u}{d t} = \frac{1}{2} \frac{\sqrt{(u + z_1 + \lambda) (u + z_2 + \lambda) (u + z_3 + \lambda)}}{\sqrt{(t + z_1) (t + z_2) (t + z_3)}}.
\end{equation}
This, together with $u(0) = 0$ and $\lim_{t \to \infty} u(t) = \infty$ finishes the proof.

The symmetric reduction from a quartic to a cubic is treated in ref.~\cite[thm.~3.2]{MR1737495}.  Suppose we have $z_1, z_2, z_3, z_4 \in \mathbb{C} \setminus (-\infty, 0)$ and take the square roots to be in the right half plane.  Then, define
\begin{equation}
    w_j = \sqrt{z_1} \sqrt{z_j} + \sqrt{z_k} \sqrt{z_l}, \qquad \{j, k, l\} = \{2, 3, 4\}.
\end{equation}
Then,
\begin{equation}
    \label{eq:symmetric_reduction}
    \int_0^\infty \frac{d u}{\prod_{i = 1}^4 \sqrt{u + z_i}} =
    \int_0^\infty \frac{d v}{\prod_{j = 2}^4 \sqrt{v + w_j^2}}.
\end{equation}
This can be shown by a change of variable
\begin{equation}
    v(u) = (\sqrt{u + z_1} \sqrt{u + z_j} + \sqrt{u + z_k} \sqrt{u + z_l})^2 - (\sqrt{z_1} \sqrt{z_j} + \sqrt{z_k} \sqrt{z_l})^2.
\end{equation}
We have
\begin{equation}
    \frac{d v}{d u} = (\sqrt{u + z_1} \sqrt{u + z_j} + \sqrt{u + z_k} \sqrt{u + z_l}) \Big(\frac{\sqrt{u + z_j}}{\sqrt{u + z_1}} + \frac{\sqrt{u + z_1}}{\sqrt{u + z_j}} + \frac{\sqrt{u + z_k}}{\sqrt{u + z_l}} + \frac{\sqrt{u + z_l}}{\sqrt{u + z_k}}\Bigr).
\end{equation}
Then,
\begin{equation}
    v + w_j^2 = (\sqrt{u + z_1} \sqrt{u + z_j} + \sqrt{u + z_k} \sqrt{u + z_l})^2
\end{equation}
and
\begin{multline}
    \sqrt{(v + w_2^2) (v + w_3^2) (v + w_4^2)} =
    (\sqrt{u + z_1} \sqrt{u + z_2} + \sqrt{u + z_3} \sqrt{u + z_4}) \\
    (\sqrt{u + z_1} \sqrt{u + z_3} + \sqrt{u + z_2} \sqrt{u + z_4})
    (\sqrt{u + z_1} \sqrt{u + z_4} + \sqrt{u + z_2} \sqrt{u + z_3}).
\end{multline}
Now the result follows by simple algebra.

If the integration domain is not $(0, \infty)$ but $(x, y)$ instead, we first make a change of variable to make the domain $(0, \infty)$.  For an integral
\begin{equation}
    \int_x^y \frac{d t}{\prod_{i = 1}^4 \sqrt{a_i + b_i t}}
\end{equation}
we make a change of variables $t = \frac{x u + y}{u + 1}$ or $u = \frac{t - y}{x - t}$.  This then yields
\begin{equation}
    \int_x^y \frac{d t}{\prod_{i = 1}^4 \sqrt{a_i + b_i t}} = \frac{y - x}{\prod_{i = 1}^4 \sqrt{a_i + b_i x}} \int_0^\infty \frac{d u}{\prod_{i = 1}^4 \sqrt{u + z_i}},
\end{equation}
where $z_i = \frac{a_i + b_i y}{a_i + b_i x}$.

Next, we can combine the domain transformation and the symmetric reduction to obtain the following result (see ref.~\cite[thm.~3.4]{MR1737495}).  For $x, y \in \mathbb{R}$ with $x > y$, and $a_i, b_i \in \mathbb{R}$ for $i = 1, 2, 3, 4$, define
\begin{gather}
    X_i = \sqrt{a_i + b_i x}, \qquad
    Y_i = \sqrt{a_i + b_i y}, \qquad i = 1, 2, 3, 4, \\
    U_{1 i} = \frac{X_1 X_j Y_k Y_l + Y_1 Y_j X_k X_l}{x - y}, \qquad i = 2, 3, 4,
\end{gather}
then
\begin{equation}
    \int_y^x \frac{d t}{\prod_{i = 1}^4 \sqrt{a_i + b_i t}} = \int_0^\infty \frac{d t}{\prod_{j = 2}^4 \sqrt{t + U_{1 j}^2}}.
\end{equation}

Indeed, we have
\begin{equation}
    \int_x^y \frac{d t}{\prod_{i = 1}^4 \sqrt{a_i + b_i t}} =
    \frac{y - x}{\prod_{i = 1}^4 \sqrt{a_i + b_i x}} \int_0^\infty \frac{d u}{\prod_{i = 1}^4 \sqrt{u + z_i}},
\end{equation}
with $z_i = \frac{a_i + b_i y}{a_i + b_i x}$.  Then, by symmetric reduction we have
\begin{equation}
    \int_0^\infty \frac{d u}{\prod_{i = 1}^4 \sqrt{u + z_i}} =
    \int_0^\infty \frac{d v}{\prod_{j = 2}^4 \sqrt{v + w_j^2}},
\end{equation}
where
\begin{multline}
    w_j = \sqrt{z_1} \sqrt{z_j} + \sqrt{z_k} \sqrt{z_l} = \\
    \sqrt{\frac{a_1 + b_1 y}{a_1 + b_1 x}} \sqrt{\frac{a_j + b_j y}{a_j + b_j x}} +
    \sqrt{\frac{a_k + b_k y}{a_k + b_k x}} \sqrt{\frac{a_l + b_l y}{a_l + b_l x}} =
    \frac{X_1 X_j Y_k Y_l + X_k X_l Y_1 Y_j}{X_1 X_j X_k X_l}.
\end{multline}
Using the rescaling $v \to \frac{v}{\alpha^2}$ for $\alpha > 0$, the identity
\begin{equation}
    \alpha \int_0^\infty \frac{d v}{\prod_{j = 2}^4 \sqrt{v + w_j^2}} = \int_0^\infty \frac{d v}{\prod_{j = 2}^4 \sqrt{v + \frac{w_j^2}{\alpha^2}}}
\end{equation}
finishes the proof.

Let us now consider the explicit case of a complete elliptic integral
\begin{equation}
    \int_{a_2}^{a_3} \frac{d t}{\sqrt{(t - a_1) (t - a_2) (a_3 - t) (a_4 - t)}}
\end{equation}
for $0 \leq a_1 \leq a_2 \leq a_3 \leq a_4$.  By a change of variables $u = \frac{t - a_2}{a_3 - t}$ we obtain
\begin{equation}
    \int_0^\infty \frac{d u}{\sqrt{u (a_2 - a_1 + (a_3 - a_1) u) (a_4 - a_2 + (a_4 - a_3) u)}}.
\end{equation}
By making a change of variables $u = \frac{a_2 - a_1}{a_3 - a_1} v$ and $u = \frac{a_4 - a_2}{a_4 - a_3} v$ we obtain, respectively
\begin{multline}
    \int_{a_2}^{a_3} \frac{d t}{\sqrt{(t - a_1) (t - a_2) (a_3 - t) (a_4 - t)}} = \\
    \frac{1}{\sqrt{(a_2 - a_1) (a_4 - a_3)}} \int_0^\infty \frac{d v}{\sqrt{v (v + 1) \Bigl(v + \frac{(a_1 - a_3) (a_4 - a_2)}{(a_1 - a_2) (a_4 - a_3)}\Bigr)}} = \\
    \frac{1}{\sqrt{(a_3 - a_1) (a_4 - a_2)}} \int_0^\infty \frac{d v}{\sqrt{v (v + 1) \Bigl(v + \frac{(a_1 - a_2) (a_4 - a_3)}{(a_1 - a_3) (a_4 - a_2)}\Bigr)}}.
\end{multline}

\acknowledgments

I am grateful to Matt von Hippel and Hjalte Frellesvig for discussions and initial collaboration.  I am also grateful to Francis Brown for a discussion about coactions.

\bibliographystyle{JHEP}
\bibliography{biblio.bib}

\end{document}